\documentclass[iop,apj]{emulateapj}
\slugcomment{Received 2019 May 8; revised 2019 June 29; accepted 2019 July 8; published 2019 August 22}

\shorttitle{COSMOS BCG Progenitors from $0.5 < z < 3.0$}
\shortauthors{Cooke et al. 2019}

\usepackage{natbib}
\usepackage{graphicx} 
\usepackage{textcomp}
\usepackage{gensymb}
\usepackage{epsfig}%

\usepackage{verbatim}
\usepackage[outdir=./]{epstopdf}
\usepackage{indentfirst}
\usepackage[colorlinks=true,citecolor=blue,linkcolor=blue]{hyperref}
\usepackage{amssymb}
\usepackage{amsmath}
\usepackage{amsfonts}%
\usepackage{fancyhdr}

\begin{document}

\title{Stellar Mass Growth of Brightest Cluster Galaxy Progenitors in COSMOS Since \lowercase{\emph{z}} $\sim$ 3}

\author{Kevin C. Cooke$^{1}$, Jeyhan S. Kartaltepe$^{1}$, K.D. Tyler$^{1}$, Behnam Darvish$^{2}$, Caitlin M. Casey$^{3}$, Olivier Le F\`evre$^{4}$, Mara Salvato$^{5,6}$, Nicholas Scoville$^{2}$}
\affil{$^1$School of Physics and Astronomy, Rochester Institute of Technology, Rochester, NY 14623, USA\\
$^2$Cahill Center for Astrophysics, California Institute of Technology, 1216 East California Boulevard, Pasadena, CA 91125, USA\\
$^3$Department of Astronomy, The University of Texas at Austin, 2515 Speedway Blvd Stop C1400, Austin, TX 78712, USA\\
$^4$Aix Marseille Universit\`e, CNRS, LAM (Laboratoire d'Astrophysique de Marseille) UMR 7326, F--13388 Marseille, France\\
$^5$MPE, Giessenbachstrasse 1, Garching D--85748, Germany\\
$^6$Cluster of Excellence, Boltzmann Strasse 2, D--85748, Germany\\
}
\email{$^1$ kcc7952@rit.edu, $^1$ jeyhan@astro.rit.edu}
\begin{abstract}
We examine the role of environment on the in situ star formation (SF) hosted by the progenitors of the most massive galaxies in the present-day universe, the brightest cluster galaxies (BCGs), from $z \sim 3$ to present in the COSMOS field.  Progenitors are selected from the COSMOS field using a stellar mass cut motivated by the evolving cumulative comoving number density of progenitors within the Illustris simulation, as well as the Millennium-II simulation and a constant comoving number density method for comparison.  We characterize each progenitor using far-ultraviolet--far-infrared observations taken from the COSMOS field and fitting stellar, dust, and active galactic nucleus components to their spectral energy distributions.  Additionally, we compare the SF rates of our progenitor sample to the local density maps of the COSMOS field to identify the effects of environment.  We find that BCG progenitors evolve in three stages, starting with an in situ SF dominated phase ($z > 2.25$).  This is followed by a phase until $z \sim 1.25$ where mass growth is driven by in situ SF and stellar mass deposited by mergers (both gas rich and poor) on the same order of magnitude independent of local environment.  Finally, at low redshift dry mergers are the dominant stellar mass generation process.   We also identify this final transition period as the time when progenitors quench, exhibiting quiescent NUV\emph{rJ} colors.
\end{abstract}


\section{Introduction}\label{sec:intro}
The evolutionary history of today's most massive galaxies is an important component in understanding the evolution of large-scale structure in the universe and galaxy populations in general, placing vital limits on the speed that baryonic matter can assemble into gravitationally bound structures.  These are massive elliptical galaxies, called brightest cluster galaxies (BCGs), and are normally located at the center of their cluster in position/velocity space \citep{Lauer:2014aa}.  They are composed of an old stellar population and a hot gaseous component in an extended envelope larger than predicted by the distribution of neighboring non-BCG ellipticals \citep{Oemler:1976aa}.  While most are quiescent, those that inhabit relaxed, massive haloes may exhibit atypically large ($>1\;M_{\odot}$ yr$^{-1}$) star formation rates (SFRs).  This is due to the surrounding intra-cluster medium (ICM) cooling and precipitating into the BCGs, forming cool-core galaxy clusters \citep[e.g.,][]{Fabian:1994aa,ODea:2008aa,ODea:2010aa,Rawle:2012aa,Fogarty:2019aa}, or due to the accretion of gas-rich satellites \citep[e.g.,][]{Tremblay:2014aa,Webb:2015ab,Cooke:2016aa}.  BCG stellar mass and SFRs may also be affected by the mass of their cluster halos at low redshift \citep[e.g.,][]{Cooke:2018aa,Gozaliasl:2018aa}.  Their atypically high mass, quiescence, and extended light profiles collectively point to a formation mechanism unique with regard to their satellites, in which most of their stars formed at $z > 2$, and conglomerated into a single BCG through many major and minor mergers \citep{De-Lucia:2007aa}. With this complex model in mind, a full identification of BCG progenitors requires the characterization of all galaxies at high redshift that are likely to merge and contribute their stellar mass onto proto-BCG ``seeds".

BCGs inhabit the high-mass end of the galaxy stellar mass function (SMF); at $z \sim 0.5$, BCGs in massive, relaxed clusters are typically $\sim 10^{11.5}\;M_{\odot}$ or greater \citep[e.g.,][]{Lidman:2012aa,Lin:2013aa,Chiu:2016aa,Cooke:2016aa}.  At this stellar mass, the population is dominated by passive (elliptical) galaxies, both BCGs and non-BCG ellipticals, that are three orders of magnitude more common than star-forming (spiral) galaxies \citep{Davidzon:2017aa}, particularly in dense environments \citep[e.g.,][]{Scoville:2013aa,Darvish:2016aa}. By starting with this low-redshift population, and then identifying the progenitor populations responsible, we examine the role of internal versus external processes at a time when proto-clusters have not finished merging into the virialized clusters observed at present day.  In situ star formation (SF) refers to any SF processes within a galaxy, such as those triggered through secular \cite[disk instability;][]{Livermore:2012aa,Livermore:2015aa} or external stimuli \citep[interactions or gas-rich mergers;][]{Sanders:1996aa,Kartaltepe:2012aa}. Ex situ stellar mass growth is the accretion of stellar mass formed outside of the galaxy, such as the old stellar populations composing the secondary galaxy in a galaxy merger \citep[e.g.,][]{Pillepich:2015aa}.

The current hierarchical model of galaxy formation relies on a combination of stellar mass growth through internal processes and mergers that build galaxies from several individual galaxies.  Mergers directly inject stellar mass into the primary galaxy concurrent with in situ SF triggered if either galaxy is gas rich.  The relative roles of internal processes and mergers have been simulated at length in large-scale cosmological models \citep[e.g., the Millennium-II Simulation (MS-II) \& the Illustris project;][]{Springel:2005aa,Vogelsberger:2014aa} and zoom-in models \citep{Narayanan:2010aa,Narayanan:2015aa,Ragone-Figueroa:2018aa}.  While there is still tension with observations of properties such as morphology \citep[e.g.,][]{Genel:2014aa}, many predictions from simulations, such as merger rates, remain within observational errors \citep{Lotz:2011aa,Rodriguez-Gomez:2017aa,Duncan:2019aa}.  The total stellar mass growth of a BCG, or any massive galaxy, can be summarized by two drivers with multiple contributors.

\begin{enumerate}
\item{In situ star formation}
\begin{enumerate}
\item{Self-sourced---a galaxy's cold gas supply condenses via local or gravitational interaction/flyby.}
\item{ICM precipitation---cooling gas from the cluster ICM precipitates down the potential well and forms stars.}
\item{Wet mergers---stars form due to the shocks and greater gas supply provided during a gas-rich merger.}
\end{enumerate}

\item{External delivery of stellar mass}
\begin{enumerate}
\item{Wet and dry mergers---immediate contribution of an existing, older stellar population.}
\item{Local interaction---capture of stellar mass from an interaction with an unbound secondary galaxy.}
\end{enumerate}
\end{enumerate}

The relative role of each of these contributors is a rich field of research, with many works specializing in one or more facet of this picture \cite[e.g.,][]{ODea:2008aa,Naab:2009aa,Loubser:2012aa,Webb:2015ab,Webb:2017aa,Contini:2018aa}.  Another factor is the role of local environment on the stellar mass assembly processes described above.  A correlation between local environment and SF is observed at low redshift \citep[e.g.,][]{Oemler:1976aa,Dressler:1980aa,Kauffmann:2004aa,Darvish:2018aa}, but is still a subject of debate at high redshift \citep[e.g.,][]{Elbaz:2007aa,Scoville:2013aa,Darvish:2016aa}.  Active galactic nucleus (AGN) activity as a function of environment at high redshift is likewise under investigation \citep[e.g.,][]{Ehlert:2013aa,Martini:2013aa,Umehata:2015aa,Alberts:2016aa,Yangyangguang:2018aa,Macuga:2019aa}. 

In this paper, we seek to understand the roles of in situ SF and external delivery toward the total growth of BCG progenitors as a function of environment and redshift.  However, identifying progenitors to specific populations in the present-day universe remains one of the most difficult, but critical, elements of galaxy evolution studies.  Due to the long timescales involved, we require a method that connects galaxies across redshift and is independent of SFR characteristics such as color.  This is often implemented as a stellar mass selection as a function of redshift and descendent stellar mass.  

A relatively straightforward method for identifying the progenitors of massive galaxies is to assume that the most massive galaxies in clusters and groups remain the most massive throughout their history \citep{van-Dokkum:2010aa}.  One implementation of this concept is the constant comoving number density method \citep{Leja:2013aa,Patel:2013aa,Dokkum:2013aa}, which assumes that the number of galaxies within a given comoving volume of the universe is constant.  As the galaxies age within the volume, the population within each stellar mass bin will grow in stellar mass with time, and inhabit the new stellar mass bin at an equal density as their old bin.  By selecting galaxies at high redshift that inhabit the universe at an equal cumulative number density, one selects the direct progenitors of the low-redshift population.  This method can be corrected to include merger partners by increasing the cumulative comoving number density cutoff (i.e., lowering the stellar mass cutoff), and this evolving cutoff has been investigated using semi-empirical approaches or by using cosmological simulations \citep[e.g.,][]{Marchesini:2014aa,Torrey:2015aa}.  

We use the evolving comoving cumulative number density selection method to identify the progenitors of BCGs out to redshift $z \sim 3$.  We select a descendant stellar mass cutoff of 10$^{11.5}\;M_{\odot}$ at $z < 0.35$ to ensure we track progenitors to BCGs in the low-redshift universe.  To characterize each progenitor, we fit the spectral energy distributions (SEDs) of each target from the far-ultraviolet (FUV) to far-infrared (FIR) to ensure sensitivity to low and obscured SFRs.  Additionally, we build upon previous studies \citep[e.g.,][]{Hill:2017aa,Torrey:2017aa} by correlating our progenitor characteristics to the local galaxy environment using redshift-sensitive adaptive kernel density maps of the COSMOS field \citep{Darvish:2015aa,Darvish:2017aa}.

  In Section~\ref{sec:inst} we discuss the data we used to construct the SEDs of our sample.   In Section~\ref{sec:methods} we describe how we implemented an evolving comoving number density selection function to acquire our BCG progenitor sample.  The median SED fits of our sample are discussed in Section~\ref{sec:analysis}.  We consider the physical implications of our results in Section~\ref{sec:disc} and we summarize our findings in Section~\ref{sec:conc}. We use the $\Lambda$CDM standard cosmological parameters of  $H_{0}$ = 70 Mpc$^{-1}$ km s$^{-1}$, $\Omega_M$ = 0.3, and $\Omega_{\mathrm{vac}}$ = 0.7 and a \citet{Chabrier:2003aa} initial mass function.

\section{COSMOS Multiwavelength Data}\label{sec:inst}
We first require a large contiguous area and volume to examine progenitors across the full range of cosmological environments.  Additionally, we need well-sampled SEDs from the FUV to the FIR to constrain the stellar mass and SFRs of the massive galaxies within these environments.  These requirements are uniquely satisfied by the COSMOS field \citep{Scoville:2007aa}.  COSMOS is a 2 sq. deg. field centered at R.A.(J2000) = 10:00:28.600, decl.(J2000) = +02:12:21.00 observed from the X-ray to the radio.  We use the public COSMOS photometric data catalog, COSMOS2015 \citep{Laigle:2016aa}, to select progenitors and construct their SEDs from the FUV to FIR.  Below, we concisely summarize each of the COSMOS datasets used in this study; for further details of COSMOS2015 see \citet{Laigle:2016aa}.

\subsection{Photometric and Spectroscopic Redshifts}

The quality of our SED fits and environment estimates depends on accurate distance measurements.  To that end, we start with COSMOS photometric redshift measurements \citep{Ilbert:2008aa} in the COSMOS2015 catalog \citep{Laigle:2016aa}.  We then crossmatch the COSMOS2015 progenitors' coordinates with the spectroscopic detections in the COSMOS spectroscopic catalog to the closest match within 1\arcsec\; of the original COSMOS2015 location.  Our BCG progenitor sample includes spectroscopic redshifts available from the 3D-\emph{Hubble Space Telescope} (HST) Survey \citep{Brammer:2012aa,Momcheva:2016aa}, Keck DEIMOS 10K Spectroscopic Survey \citep{Hasinger:2018aa}, FMOS-COSMOS Survey \citep{Kashino:2013aa,Kartaltepe:2015ab,Silverman:2015aa}, the Gemini GMOS-S spectra of \citet{Balogh:2011aa}, COSMOS AGN Spectroscopic Survey \citep{Trump:2007aa,Trump:2009aa}, the Keck LRIS spectra of \citet{Casey:2017aa}, MOIRCS Deep Survey \citep{Yoshikawa:2010aa}, the Keck MOSFIRE spectra of \citet{Trakhtenbrot:2016aa}, the Keck DEIMOS spectra of \citet{Capak:2011aa} and \citet{Mobasher:2016aa}, the SINFONI spectra of \citet{Perna:2015aa}, the zCOSMOS Survey \citep{Lilly:2007aa}, the VIMOS Ultra-Deep Survey \citep{Le-Fevre:2015aa}, and the HST/WFC3 grim spectra of \citet{Krogager:2014aa}.

		\subsection{GALEX}
		We use FUV and near-ultraviolet (NUV) band point-spreadfunction (PSF)-fit photometric magnitudes from the \emph{Galaxy Evolution Explorer} \citep[\emph{GALEX};][]{Martin:2005aa} to constrain the degree of unobscured SF.  \emph{GALEX} observations in the COSMOS2015 catalog were originally reduced by \citet{Zamojski:2007aa}.  We correct the \emph{GALEX} FUV-\emph{Spitzer} IRAC4 observations for Milky Way foreground extinction using a galactic reddening of $R_v$ = 3.1 \citep{Morrissey:2007aa} and \emph{E(\bv)} values from the dust maps of \citet{Schlegel:1998aa}. 
	
		\subsection{Canada--France--Hawaii Telescope}
Canada-France-Hawaii Telescope (CFHT)/MegaPrime \citep{Aune:2003aa,Boulade:2003aa} $u^{*}$ observations of the COSMOS field were taken with a consistent depth of m$_{u^{*}}$ $\sim$ 26.4 \citep{Capak:2007aa}.   
		
		\subsection{Subaru}
		To constrain our stellar mass estimates, we need sensitive optical continuum measurements across as many bands as possible.   Therefore, we include Subaru/Suprime-Cam optical observations using five broadband filters (\emph{B, V, R}, $i+$, $z++$) and 11 medium-band filters (IA427, IA464, IA484, IA505, IA527, IA574, IA624, IA679, IA738, IA767, and IA827), observed to a $3\sigma$ depth of $m_{AB}\sim25.2$ or deeper.  Out of these filters, IA464 suffers the worst resolution with a PSF FWHM of 1.89" \citep{Taniguchi:2007aa,Taniguchi:2015aa}.  
		
		\subsection{Vista}
		We include Vista/VIRCAM \citep{Sutherland:2015aa} \emph{J-, H-}, and \emph{K}-band observations from the UltraVISTA-DR2 survey \citep{McCracken:2012aa}. These have limiting magnitudes $m_{AB}$ of  24.7, 24.3, and 24.0 respectively.
		
		\subsection{\emph{Spitzer} Observations}
		 Further IR observations of the COSMOS field are available from the SPLASH \citep{Steinhardt:2014aa} and S-COSMOS surveys \citep{Sanders:2007aa} using the \emph{Spitzer Space Telescope} \citep[\textit{Spitzer};][]{Werner:2004aa}.  Data from \textit{Spitzer}'s Infrared Array Camera (IRAC)'s  3.6, 4.5, 5.7, and 8.0 \micron\; channels \citep[for more information, see][]{Fazio:2004aa} have PSF widths 1\farcs6, 1\farcs6, 1\farcs8, and 1\farcs9, respectively, and were observed down to a 3$\sigma$ depth of $m_{AB}$ of 25.5, 25.5, 23.0, and 22.9, respectively.  We also use MIPS \citep{Rieke:2004aa} 24 $\micron$ measurements, which act as an important constraint to the total IR fit.  MIPS 24 $\micron$ was observed to a 5$\sigma$ depth of 71 $\mu$Jy.
		
		\subsection{\emph{Herschel} Observations}
		To accurately estimate the obscured AGN and SF activity in our sample, we use FIR observations to constrain the FIR peak.  COSMOS FIR observations include the \emph{Herschel Space Observatory} \cite[\emph{Herschel};][]{Pilbratt:2010aa} Photoconductor Array Camera and Spectrometer \citep[PACS; ][]{Poglitsch:2010aa} 100 and 160 \micron\; bands and Spectral and Photometric Imaging Receiver (SPIRE) 250, 350, and 500 \micron\; bands.  The PACS Evolutionary Probe \citep{Lutz:2011aa} data used in this study were observed down to a 3$\sigma$ depth of 5 and 10.2 mJy for 100 and 160 \micron\; bands, respectively.  \emph{Herschel} Multi-tiered Extragalactic Survey \citep{Oliver:2012aa} SPIRE 250 \micron, 350 \micron, and 500\micron\; observations reach a 3$\sigma$ depth of 8.1, 10.7 mJy, 15.4 mJy, respectively.  
		
		Blending is possible at the longest wavelengths due to the large beam size (18.1\arcsec\; at 250 \micron, 24.9\arcsec\; at 350 \micron, 36.6\arcsec\; at 500 \micron), therefore de-blending was performed band by band using the next most-resolved observation starting with \emph{Spitzer} IRAC observations as priors \citep{Lee:2010aa}. Blending is the largest issue at 500 \micron, so to ensure that the 500 $\micron$\; detections correspond to our progenitor sample, we use the `clean index' from \citet{Elbaz:2011aa}.  In brief, a progenitor's 500 $\micron$ data are rejected if there is more than one bright neighbor within 1.1x the FWHM at 24 $\micron$ or there are bright neighbors within 1.1$\times$ the FWHM of the respective \emph{Herschel} FIR band, with a bright neighbor defined as $S_{\mathrm{Neighbor}}$/$S_{\mathrm{Target}} >$ 0.5.  

\section{Methods}\label{sec:methods}
\subsection{Cumulative Number Density Selection Function}\label{sec:numberdensity}
\begin{figure}
\begin{center}
\begin{tabular}{c}
\includegraphics[width=0.49\textwidth]{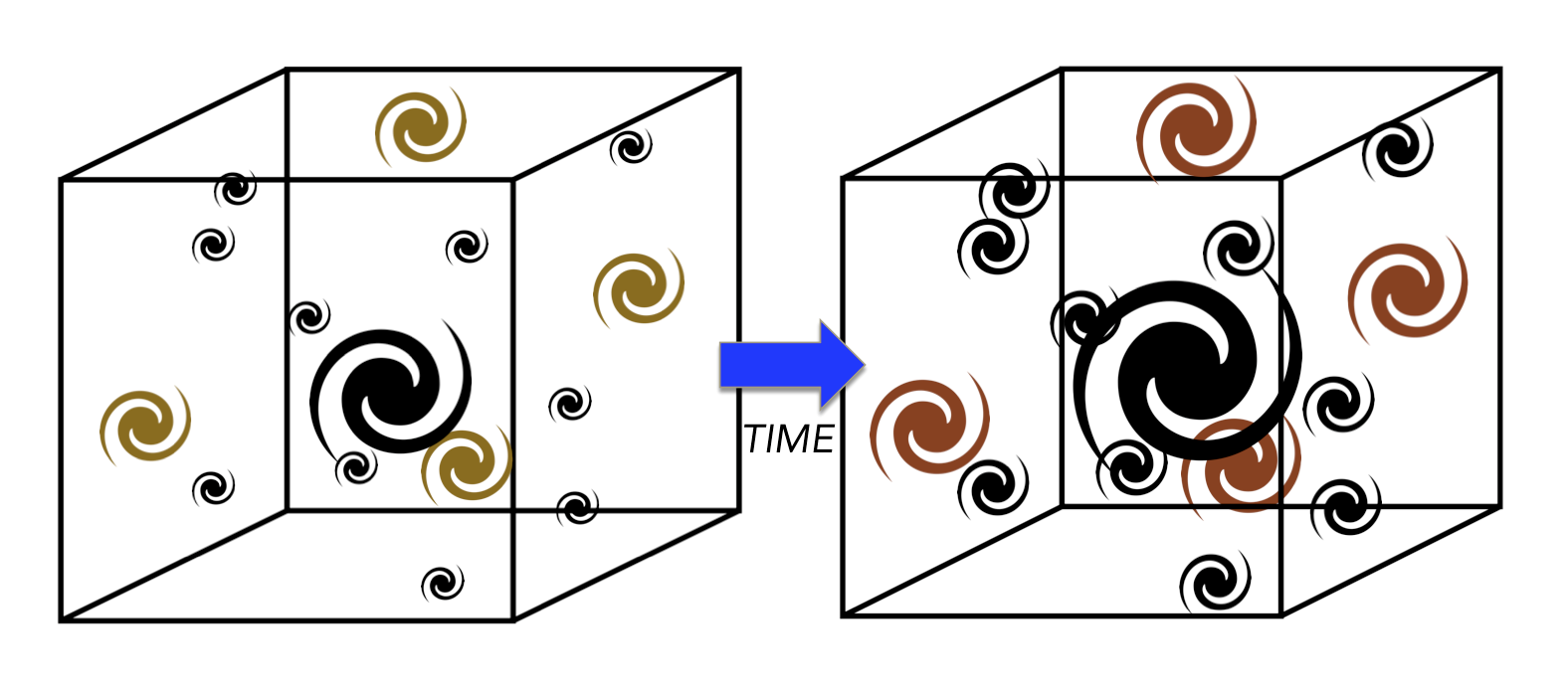}
\end{tabular}
\end{center}
\caption
{ \label{fig:boxexample} Example of the constant number density selection method, using the assumption of a constant number of galaxies in the universe (total box).  A population of low-redshift, quiescent galaxies (shaded red spirals) on the right correspond to their high-redshift progenitors on the left (shaded yellow spirals) by the density they inhabit the universe, irrespective of parameters such as color or star formation activity.}
\end{figure}

Due to the long timescales involved in an individual galaxy's evolution, large samples over many epochs are needed to probe the growth of stellar mass in the universe.  Selecting a population of progenitors at high redshift is difficult because the entire point of the exercise is that an important characteristic of the progenitor population remains unknown. The constant comoving cumulative number density method introduced by \citet{van-Dokkum:2010aa} was a solution to this problem.  This method assumes a constant number of galaxies in the past and present universe.  To identify the progenitors of massive galaxies in the present day, one would measure the density they inhabit in a volume element of the present-day universe and identify a corresponding population at high redshift that occupied the universe at the same volume density (illustrated in Figure\;\ref{fig:boxexample}).

This method has been successfully used to identify some of the progenitors for today's population of massive galaxies \citep{van-Dokkum:2010aa,Patel:2013aa,Morishita:2015aa}, but it is biased toward higher-mass progenitors.  This bias arises because the galaxies that merge onto the selected progenitor galaxy are not included.  Secondary galaxies are often lower mass (of higher number density) and therefore their exclusion results in a bias toward lower number density and higher-mass progenitors \citep{Mundy:2015aa}.   Progenitors that evolved via a merger-dominant evolutionary path would not be selected, as these low-mass progenitors would inhabit a lower stellar mass (and higher number density) parameter space than the rest of the progenitor population.  The disparity between predicted and `actual' progenitor stellar masses could reach a factor of two at high redshift \citep{Leja:2013aa,Mundy:2015aa}.  To better correct for the effect of mergers, an evolving number density method was developed by tracking the median progenitor mass in cosmological simulations \citep{Mundy:2015aa,Torrey:2015aa,Jaacks:2016aa}.

An evolving number density selection is derived by identifying a population in a low-redshift slice of the universe and tracing back in redshift the number density that corresponds to the median stellar mass of their most massive progenitors.  This method has been used with multiple simulations, and our analysis includes the Illustris simulation number density evolution tracks \citep{Torrey:2017aa, Torrey:2015aa}, with comparison tracks from MS-II \citep{Boylan-Kolchin:2009aa} and a constant number density cut as discussed in Section~\ref{sec:disc}.

The Illustris simulation is a cosmological hydrodynamic simulation of volume 1.2079 $\times$10$^6$ Mpc$^3$ that includes dark and baryonic physics \citep{Vogelsberger:2014aa}.  Baryonic processes include radiative cooling, SF and supernova feedback, and AGN feedback.  Baryonic mass is resolved with particles of mass 1.3 $\times$ 10$^6$ $M_{\odot}$.  The inclusion of baryonic physics enables Illustris to more accurately model the evolution of SF at late times in the universe as AGN and outflows heat and push out the gas supply within each galaxy. 

To estimate stellar masses of progenitors observed in COSMOS, we use \citet{Davidzon:2017aa}'s SMF to identify the cumulative number density of galaxies with stellar masses above 10$^{11.5}$ $M_{\odot}$ below $z \sim 0.35$, which corresponds to the regime of low-redshift BCGs \citep[e.g.,][]{Lidman:2012aa,Liu:2012aa,Fraser-McKelvie:2014ab,Cooke:2016aa}.  This high stellar mass selection specifically examines high-mass BCGs commonly hosted in clusters with total mass $> 10^{14} M_{\odot}$ \citep[e.g.,][]{Cavagnolo:2009aa,Postman:2012aa,Lin:2013ab} to better understand how the most massive galaxies acquired their stellar populations. We choose to use the SMF of the total population and not only the passive population, as the latter outnumbers the active population by a factor of ten at this mass and redshift and we do not wish to bias our selection toward the most passive progenitors.  The results shown in Section \ref{sec:disc} are consistent within errors if we exclude or include active descendants at low redshift. Once the density of the total descendant population is identified, we determine the cumulative comoving number density of BCG progenitors in redshift slices out to $z \sim 3$ using an Illustris-derived number density evolution relation \citep[Equation (A4) of][]{Torrey:2017aa},
\begin{displaymath}
  \langle N(z) \rangle' = N_0 + \Delta z(A'_{0} + A'_{1} N_{0} + A'_{2} N^{2}_{0})
  \end{displaymath}
  \vspace{-10pt}
 \begin{displaymath}
 \;\;\;\;\;\;\;\; + \Delta z^2(B'_{0} + B'_{1} N_{0} + B'_{2} N^{2}_{0}),
\end{displaymath}
 where $N_0$ represents the cumulative comoving number density of BCGs below $z \sim 0.35$, and the constants $A'_{0,1,2}$ and $B'_{0,1,2}$ are best-fit parameters to the backward median number density evolution.

\subsection{Application to Observations}
\begin{figure*}
\begin{center}
\begin{tabular}{c}
\includegraphics[height=9cm]{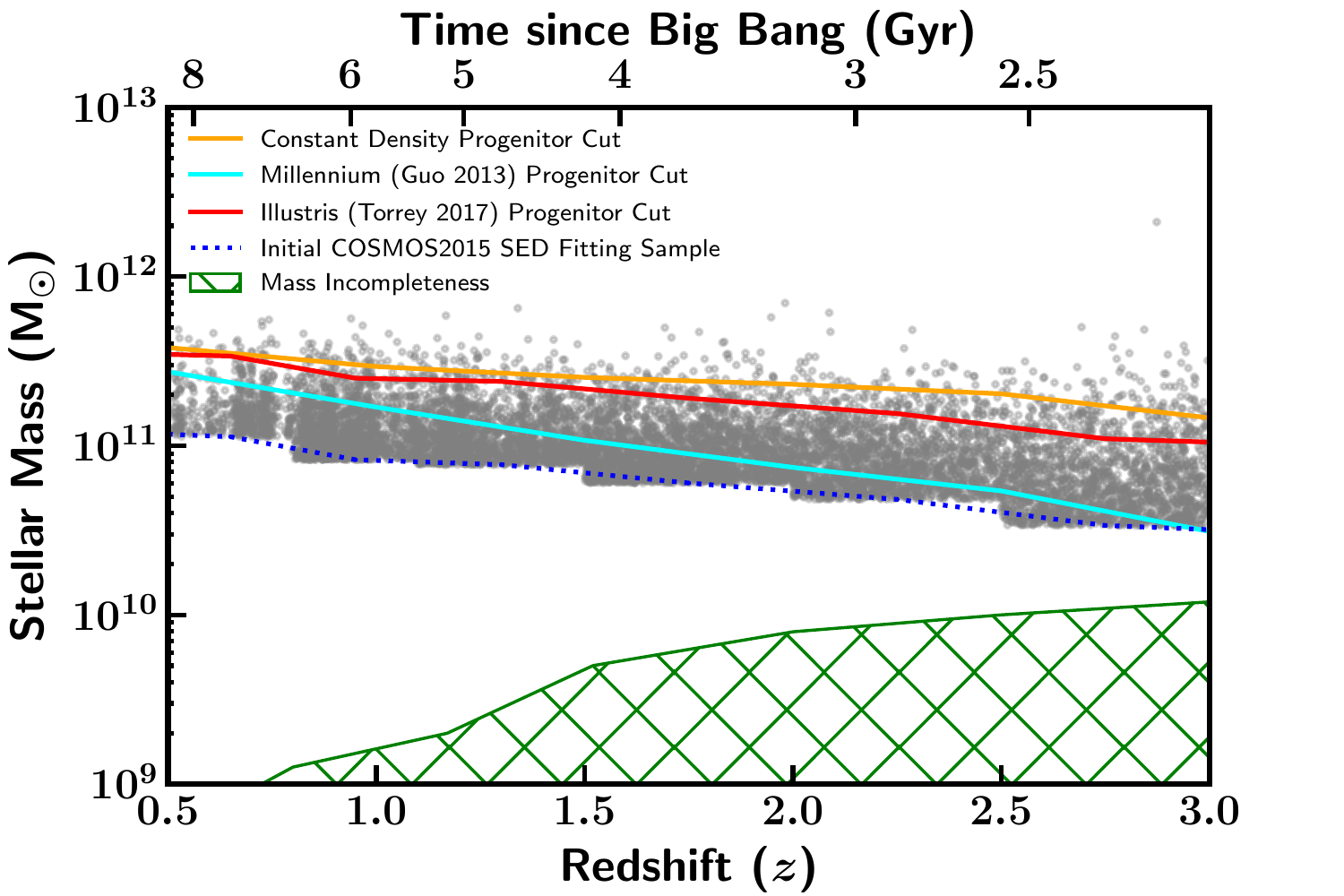}
\end{tabular}
\end{center}
\caption
{ \label{fig:evolutiontracker} Stellar mass selection function of our sample via constant comoving number density (orange), and the evolving comoving number density predicted via Illustris (red) and MS-II (teal).  We also identify the stellar mass cutoff for the massive galaxy population that deposits 50\% of the merger delivered stellar mass at that redshift (blue dotted) and plot the galaxies from COSMOS2015 that satisfy this lower mass cut (gray circles). This lower cut enables us to identify galaxies with initial COSMOS estimates below our Illustris cutoff that may be refit to higher masses.  The mass incompleteness limit is hatched green \citep{Laigle:2016aa}, and is defined as the stellar mass above which 90\% of galaxies are detected given the $K_s$ limiting magnitude of the COSMOS field observations \citep{Pozzetti_2010,Davidzon:2013aa,Ilbert:2013aa,Moustakas:2013aa}.}
\end{figure*}

To construct our final sample, we take the cumulative comoving number density predictions estimated above and calculate the stellar mass corresponding to the progenitor cumulative number density in that redshift slice (red line in Figure~\ref{fig:evolutiontracker}) given the observed COSMOS SMF \citep{Davidzon:2017aa}.  To maintain sample completeness, we fit the SEDs of an expanded sample down to a lower stellar mass limit than required by our Illustris-derived selection function.  By fitting COSMOS galaxies within a factor of 3 of the Illustris selection function, we hope to recover any galaxies with underestimated masses in their initial fit in the COSMOS2015 catalog. After fitting our initial sample of prospective progenitors (dotted blue line in Figure~\ref{fig:evolutiontracker}), we select galaxies that have a fit stellar mass (described in Section~\ref{sec:analysis}) above the Illustris evolving comoving number density method mass cut for that given redshift bin (red line in Figure~\ref{fig:evolutiontracker}).  Our final sample includes 1444 BCG progenitors from $0.5 < z < 3.0$. 

\begin{deluxetable*}{lccccccccr}
\tabletypesize{\footnotesize}
\tablecolumns{9} 
 \tablecaption{Illustris-selected BCG Progenitor Sample Distribution
 \label{tab:sample}}
 \tablehead{\colhead{$z$} & \colhead{Total Progenitors}& \colhead{Total Photo-$z$} & \colhead{Total Spec-$z$} & \colhead{Total AGN$^\dagger$}& \colhead{X-ray AGN} & \colhead{MIR AGN} & \colhead{Radio AGN} & \colhead{SED3FIT AGN$^\ddagger$} } 
 \startdata 
 0.5-0.8 & 26 & 2 & 24 & 2 (7.7\%)& 1 & 1 & 2 & 0 &\\ 
 0.8-1.1 & 180 & 74 & 106 & 14 (7.8\%) &1 2 & 6 & 0 & 5 &\\ 
 1.1-1.5 & 216 & 156 & 60 & 8 (3.7\%) & 4 & 4 & 1 & 3 &\\ 
 1.5-2.0 & 401 & 371 & 30 & 29 (7.2\%) & 22 & 15 & 1 & 12 &\\ 
 2.0-2.5 & 347 & 308 & 41 & 32 (9.2\%) & 15 & 24 & 1 & 9 &\\ 
 2.5-3.0 & 274 & 254 & 20 & 31 (11.3\%) & 10 & 26 & 0 & 8 &
 \enddata
 \tablecomments{Galaxies selected from the COSMOS2015 catalog using the stellar mass cut predicted via the median BCG progenitor population in the Illustris simulation. $^\dagger$ Total AGN fractions represent the fraction of the AGN population identified through any method out of the total progenitor sample in that redshift bin. An AGN may satisfy multiple criteria, and are counted in any classification count that they satisfy.  $^\ddagger$Defined as AGN by a fractional residual $>$ 0.4 at 8.0 \micron\;(\emph{Spitzer} IRAC4) after fitting with a MAGPHYS two-component model (see Section \ref{sec:analysis}).}
\end{deluxetable*}

\subsection{Comparison Sample Selection Methods}
We also compare our results based on the Illustris selection function to other commonly used selection methods.  The MS-II simulation \citep{Boylan-Kolchin:2009aa} is a dark-matter-only simulation and does not include baryonic physics.  Number density is thus derived from dark matter halos characteristic of our low-redshift massive elliptical sample.  For MS-II halos, we identify the stellar mass of a given halo mass' s baryonic counterpart using \citet{Guo:2013aa} Millennium WMAP 7 (MR7) semi-analytic galaxy models scaled to the MS-II simulation.  We start similarly to our Illustris method by using \citet{Davidzon:2017aa}'s SMF to identify the cumulative number density of galaxies with stellar masses above 10$^{11.5}$ M$_{\odot}$ at $z \sim 0.35$.  Then we identify the dark matter halo mass corresponding to this number density cutoff.  We identify the most massive halo progenitor in each higher-redshift bin, and find the number density corresponding to the median of the most massive halo.  This number density is then used with the corresponding \citet{Davidzon:2017aa} SMF at that redshift to identify the galaxy progenitor population from \citet{Guo:2013aa}'s Millennium-II mock galaxy catalog.  We make the assumption that a single BCG progenitor lies in each halo.  The MS-II derived evolution track is shown as the teal solid line in Figure~\ref{fig:evolutiontracker}.  We discuss how the MS-II-selected sample results compare to our primary Illustris-selected results in Section~\ref{sec:millcompar}.

Finally, we wish to compare our results to those derived from the original constant number density method.  Our constant number density comparison sample is chosen by identifying the cumulative comoving number density of ellipticals with stellar masses 10$^{11.5}$ $M_{\odot}$ at $z \sim 0.35$ in \citet{Davidzon:2017aa}'s SMF, and identifying the stellar mass of the cumulative population above that density in all redshift bins out to $z \sim 3$.  This stellar mass and redshift combination corresponds to the massive BCG population in the low- to intermediate-redshift universe \citep[e.g.,][]{Lidman:2012aa,Liu:2012aa,Fraser-McKelvie:2014ab,Cooke:2016aa}. Our constant number density evolution track is shown as the orange solid line in Figure~\ref{fig:evolutiontracker}.  A comparison between the constant number density selection method and the evolving number density method used for the main results of this work is included in Section~\ref{sec:constantcompar}.

\begin{figure*}[t]
\begin{center}
\begin{tabular}{c}

\includegraphics[width=0.8\textwidth]{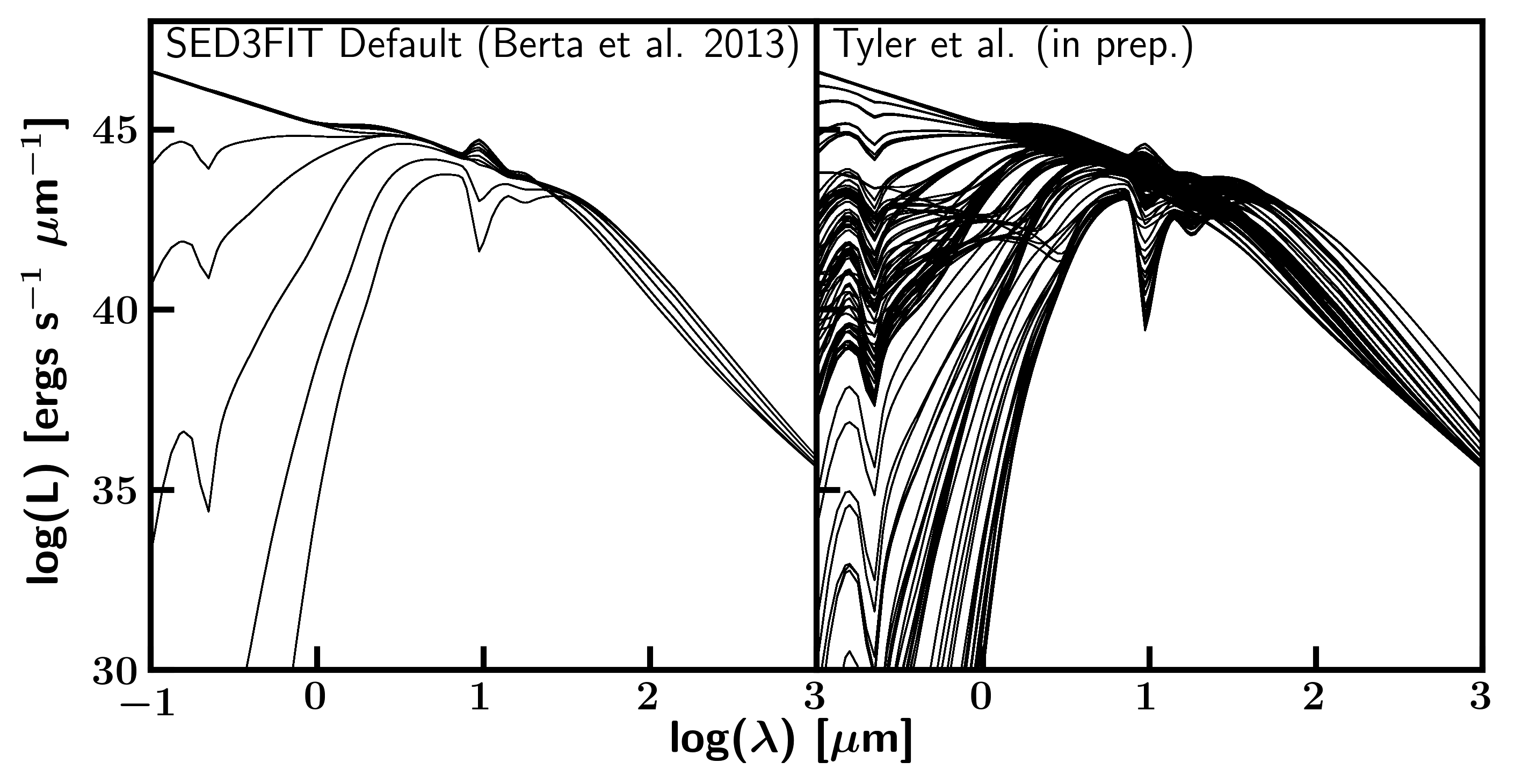}
\end{tabular}
\end{center}
\caption{Left: The original 10 \citet{Feltre:2012aa} AGN models included in SED3FIT with viewing angles of 0\textdegree or 90\textdegree. Right:  our selection of 240 AGN torus+disk models from \citet{Feltre:2012aa} equally sampling viewing angle.}
\label{fig:AGNmodels}
\end{figure*} 

\subsection{Quantifying Local Environment}\label{sec:map match}
With progenitor stellar mass selection criteria calculated, our next step is to select galaxies likely to merge onto the BCG `seeds' over time given their local environment.  The sample selection discussed in Section~\ref{sec:numberdensity} identified progenitors, irrespective of environment, that are predicted to relocate to the overdense regions where present day BCGs ultimately reside. Therefore we use density maps of the COSMOS field to identify progenitors already in overdense regions (very likely to merge into a BCG within a Hubble time) and underdense regions (less likely to merge into a BCG).  

We quantify local environment using density field maps of the COSMOS field produced using the `weighted' adaptive kernel smoothing approach of \citet{Darvish:2015aa}.  The density field is evaluated in a series of overlapping redshift slices. \citet{Darvish:2015aa} assign galaxy weights that are proportional to the likelihood of a galaxy belonging to a redshift. They are estimated by measuring what fraction of a galaxy's redshift distribution is within a particular redshift slice. Given these weights, the density field is then adaptively smoothed with a Gaussian kernel whose `global' width is 0.5 Mpc. The adaptive widths are evaluated around this global width, depending on how sparse or dense the local neighborhood of each galaxy is \citep[see][for details]{Darvish:2015aa}.

\subsection{Identification of AGN}\label{sec:AGNID}
We also quantify the contribution of the AGN emission to the IR regime of progenitor SEDs.  This step is important, as a dominant AGN could cause the SED fit solution to overestimate stellar mass, and an IR-bright AGN could result in a fit with an overestimated SFR.  We crossmatch our sample with the \emph{Chandra} X-ray observations of the COSMOS field \citep{Elvis:2009aa,Civano:2016aa,Marchesi:2016aa,Lanzuisi:2017aa} to identify X-ray-bright AGN.  We also use the \emph{Spitzer} IRAC photometry available for the COSMOS field to identify obscured AGN through their mid-infrared (MIR) colors using the \citet{Donley:2012aa} IRAC color criteria. Radio AGN are identified by crossmatching our progenitor sample with the radio AGN identified in the VLA-COSMOS 3 GHz Large Project \citep{Delvecchio:2017aa,Smolcic:2017aa,Smolcic:2008aa}.  Our full progenitor and AGN sample selection totals and statistics are listed in Table~\ref{tab:sample}.

\section{SED Fitting}\label{sec:analysis}
\subsection{Software and Methodology}\label{sec:sedfitting}
To determine the SFR and total stellar mass of each progenitor, we require a way to reliably fit thousands of galaxy SEDs from rest-frame FUV to FIR.  To do so, we choose the SED fitting software MAGPHYS \citep{da-Cunha:2008aa} with 50,000 stellar population models generated from \cite{Bruzual:2003aa} and 50,000 infrared dust models from \citet{da-Cunha:2008aa}. 

MAGPHYS fits the SED of each galaxy in two stages.  First, it estimates the best-fit stellar model between $0.0912\micron \lesssim \lambda \lesssim 10\micron$  from the \cite{Bruzual:2003aa} stellar population models while taking into account dust obscuration.  Once the optical component is fit, MAGPHYS fits the infrared from 2.5 to 1000 \micron\; while taking the total luminosity of obscured starlight found during the optical fitting as a prior.  FIR characteristics are calculated between 8 and 1000 \micron. This ensures energy balance between the energy absorbed and emitted by the dust component.  

\begin{figure*}
\begin{center}
\begin{tabular}{c}
\includegraphics[width=0.9\textwidth]{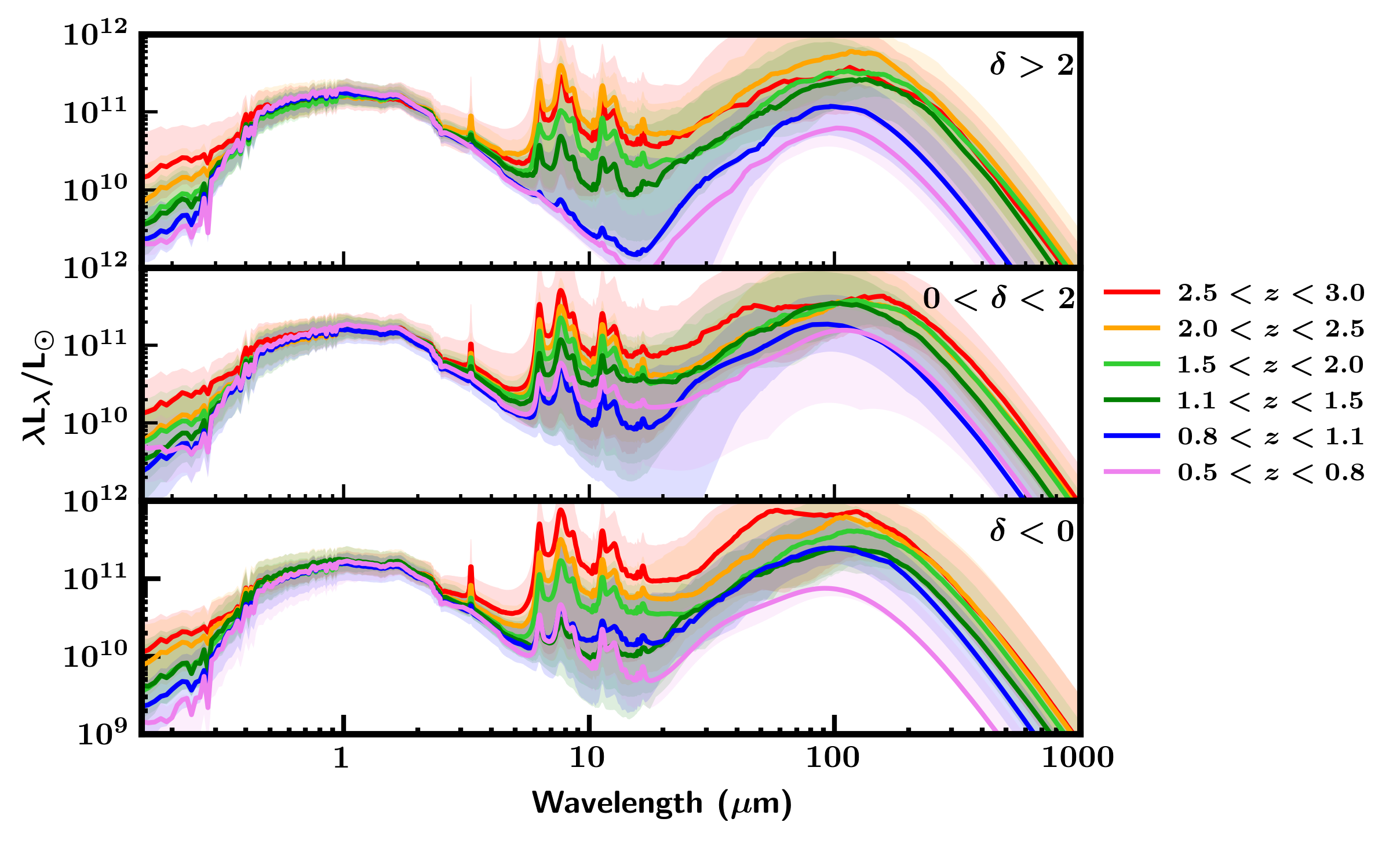}
\end{tabular}
\end{center}
\caption
{ \label{fig:medianSEDs} Median spectral energy distributions (solid lines) with $1\sigma$ confidence intervals (shaded regions) of massive progenitors between $0.5 < z < 3$ for dense regions (top), intergroup regions (middle), and field regions (bottom).  Dense environments are defined as $\frac{\delta}{\delta_{\mathrm{median}}}  >$ 2, where $\delta_{\mathrm{median}}$ is the density at that location and $\delta_{\mathrm{median}}$ represents the median density at that redshift.  Intergroup is defined as $1 < \frac{\delta}{\delta_{\mathrm{median}}}  < 2$.  Field environments are densities $\frac{\delta}{\delta_{\mathrm{median}}}  < 1$.  Color coding corresponds to redshift bins and is identical in all three subplots. We find that across all environments, FIR luminosity decreases with decreasing redshift.  However, MIR luminosity decreases with decreasing redshift at a much higher rate in dense environments in comparison to intergroup or field starting in our $0.8 < z < 1.1$ redshift bin.  This may be due to either a lack of warm dust radiating in the MIR or a lack of SFR illuminating the dust supply in progenitors residing in dense environments.  The latter is more likely as the SFR is also decreasing with decreasing redshift.}
\end{figure*}

At time of writing, the public version of MAGPHYS does not include an AGN component.  This is an important consideration, as additional luminosity across the optical and near-IR from an AGN accretion disk and the dusty torus surrounding it may skew stellar mass estimates to higher values.  Additionally, any UV or IR emission may also cause SFRs to be overestimated.  Therefore to identify potential overestimations due to a lack of an AGN fitting component,  we investigated several goodness-of-fit metrics to identify which progenitors require an AGN component.  We find that a fractional difference cutoff of 0.4 between the MAGPHYS fit 8.0 $\micron$ (\emph{Spitzer} IRAC4) flux and the observed value worked most consistently to identify sources that require an AGN component in the MIR (for further details, see K.D. Tyler et al. 2019, in preparation).  This parameter measures how poorly the stellar and dust models fit the slope of the MIR, which could indicate the requirement of a third (AGN) component.  We find that $\sim$1\% of our progenitor sample fits have MAGPHYS fractional 8.0 $\micron$ residuals $> 0.4$, and we refit this subsample using the package SED3FIT.  By choosing to select and fit AGN components to only the galaxies that have poor IR fits without an AGN, we prevent the overfitting of our sample and the possible underestimation of SFRs that an overestimation of AGN activity and frequency would induce.

SED3FIT \citep{Berta:2013aa} works similarly to MAGPHYS by fitting a series of optical models first, and then using the optical results as a prior during the fitting of the IR component.  SED3FIT, however, includes AGN torus and accretion disk emission \citep{Feltre:2012aa} across both components while maintaining energy balance.  For the IR-bright AGN identified through their $8\mu$m residuals, we use the SFR and stellar mass estimates from SED3FIT instead of MAGPHYS in the following results.

\subsection{AGN Template Library}
The original AGN template library included in the public SED3FIT distribution included 10 AGN templates from the \citet{Feltre:2012aa} AGN template library.  These templates were selected at extreme viewing angles, effectively probing archetypical Type-1 and Type-2 AGN.  These models assume either negligible or strong extinction with respect to line of sight.  While satisfactory for first-order fits, the model of the obscuring torus surrounding the supermassive black hole in an AGN has changed in recent years to a clumpy torus with regions of high and low obscuration at high and low viewing angles \citep[e.g.,][]{Krolik:1988aa,Shi:2006aa,Markowitz:2014aa}.  Therefore, we expand our AGN template library to include 240 models (K.D. Tyler et al. 2019, in preparation) equally probing between 0$\degree$ and 90$\degree$.  Shown in Figure~\ref{fig:AGNmodels}, this new library contains models that fit AGN at intermediate obscurations and viewing angles.

In Figure~\ref{fig:medianSEDs}, we plot the median SED of the progenitors found in three environments in six redshift bins with $1\sigma$ confidence intervals (shaded regions).  We classify each environment as dense \big($\frac{\delta}{\delta_{median}}  >$ 2\big), intergroup \big(1 $<$ $\frac{\delta}{\delta_{median}}  <$ 2\big), or field \big($\frac{\delta}{\delta_{median}}  <$ 1\big).  In the top panel, we find that progenitors in dense regions have SEDs similar to those in other environments at high redshift ($z > 1.1$), but have $\sim\frac{1}{5}$ the MIR emission (10--20$\mu$m) at $z < 1.1$ in comparison to progenitors at lower density at low redshift.  This indicates a divergence in evolution due to environment at low redshift, where massive progenitors in high-density environments undergo a more efficient removal of their warm gas and dust supplies.

\section{Results and Discussion}\label{sec:disc}
To characterize our total sample of 1444 BCG progenitors, we use the stellar parameters fit by MAGPHYS for 1407 progenitors and the estimates from SED3FIT for 37 progenitors due to their significant 8.0 $\micron$ residuals in their original MAGPHYS fits (hereafter referred to as SED3FIT AGN).  The SED3FIT AGN fraction is $<1\%$ in any redshift bin, with an overall AGN fraction found through X-ray, MIR, and radio methods of $\sim8\%$ (Table \ref{tab:sample}).  For comparison, previous Sunyaev-Zeldovich-selected studies have found X-ray cavities indicating past AGN activity in $\sim$7$\%$ of low-redshift BCGs \citep{Hlavacek-Larrondo:2015aa}.  

\subsection{SF and Stellar Mass Estimates}
\begin{figure}
\begin{center}
\begin{tabular}{c}
\includegraphics[width=0.49\textwidth]{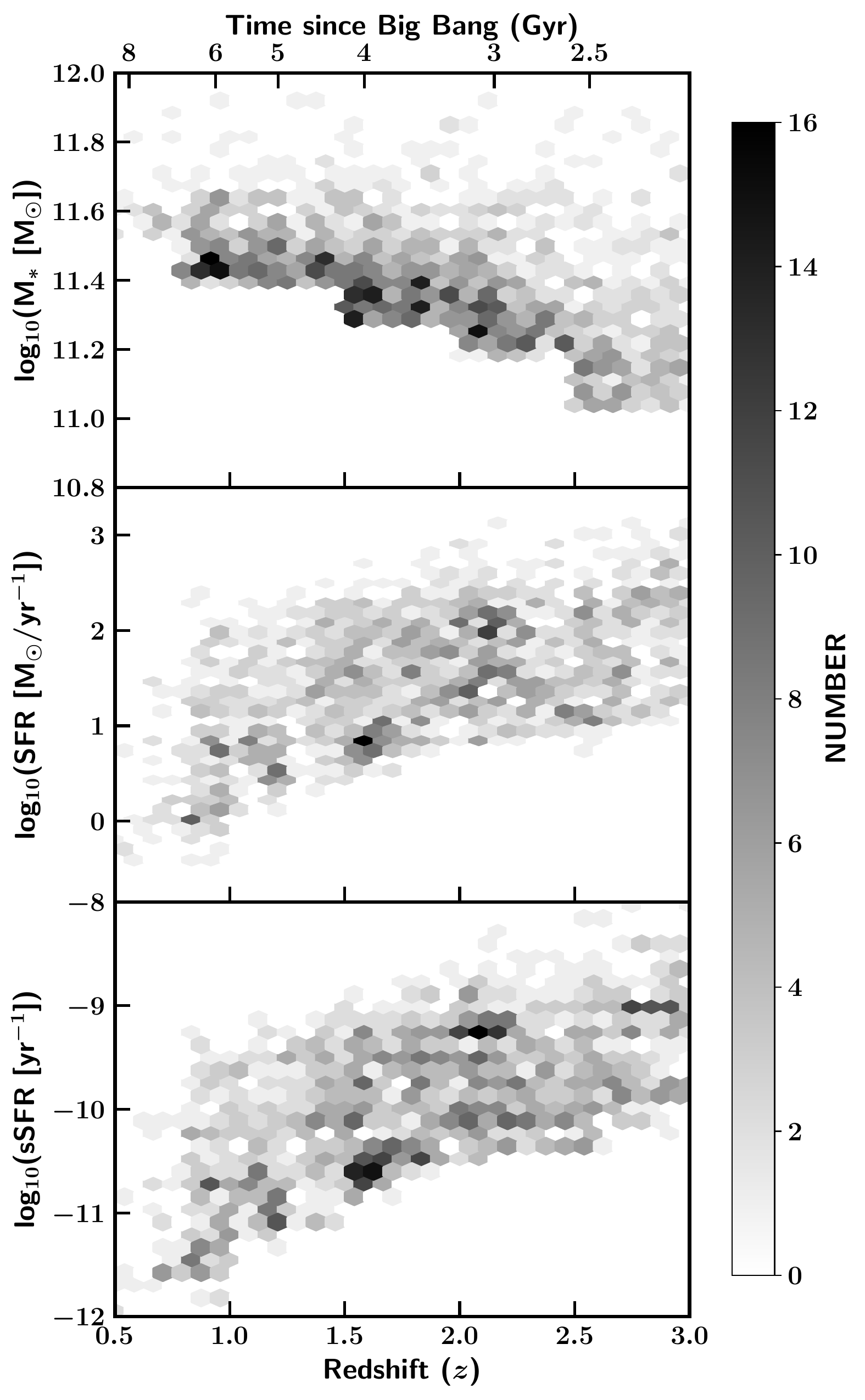}
\end{tabular}
\end{center}
\caption
{ \label{fig:sfrvsz} Top: stellar mass vs. redshift of massive BCG progenitors between $0.5 < z < 3$ fit using MAGPHYS and SED3FIT.  This plot includes all sources with stellar mass estimates above the stellar mass selection cut described in Section~\ref{sec:methods}.  Middle: SFR versus redshift of massive BCG progenitors between $0.5 < z < 3$. Bottom: specific SFR versus redshift of massive BCG progenitors between $0.5 < z < 3$.}
\end{figure}

Our sample spans an order of magnitude in stellar mass from 10$^{11}$ -- 10$^{12}$ $M_{\odot}$ at $z \sim 3$ (Figure~\ref{fig:sfrvsz}), motivated by the stellar mass selection function shown in Figure~\ref{fig:evolutiontracker}.  Our sample below $z \sim 0.8$ is subject to small-number statistics due to the low volume available in COSMOS at this distance ($\sim$ 4.4 $\times$ 10$^6$ Mpc$^3$).  

Shown in Figure~\ref{fig:sfrvsz}, both SFR (averaged over 10$^{8}$ yr) and specific SFR (sSFR) span three orders of magnitude from $0.5 < z < 3$.  Progenitors do not span this parameter space equally, with a concentration of highly star forming progenitors with sSFR $\sim$ 10$^{-9.25}$ yr$^{-1}$ at $z > 2$, that transition to a lower sSFR of 10$^{-10.75}$ yr$^{-1}$ at $z < 2$. This is partly driven by our stellar mass selection function, which is limiting the stellar mass parameter space to higher stellar masses at low redshift.  Our observed relation at the bottom of Figure~\ref{fig:sfrvsz} is systematically an order of magnitude below the sSFR--$z$ correlation found for less-massive (10$^{10.5}$ $M_{\odot}$) galaxies found in the COSMOS field \citep{Davidzon:2018aa}, consistent with galactic `downsizing' where less-massive galaxies continue to form stars while massive galaxies are quenching \citep{Cowie:1996aa,Perez-Gonzalez:2008aa}.  This will yield descendants dominated by the old stellar populations, just as observed in BCGs in the nearby universe \citep{Loubser:2009aa}.

\subsection{Evolution of Stellar Mass Growth}
\begin{figure*}[t]
\begin{center}
\begin{tabular}{c}
\includegraphics[height=9cm]{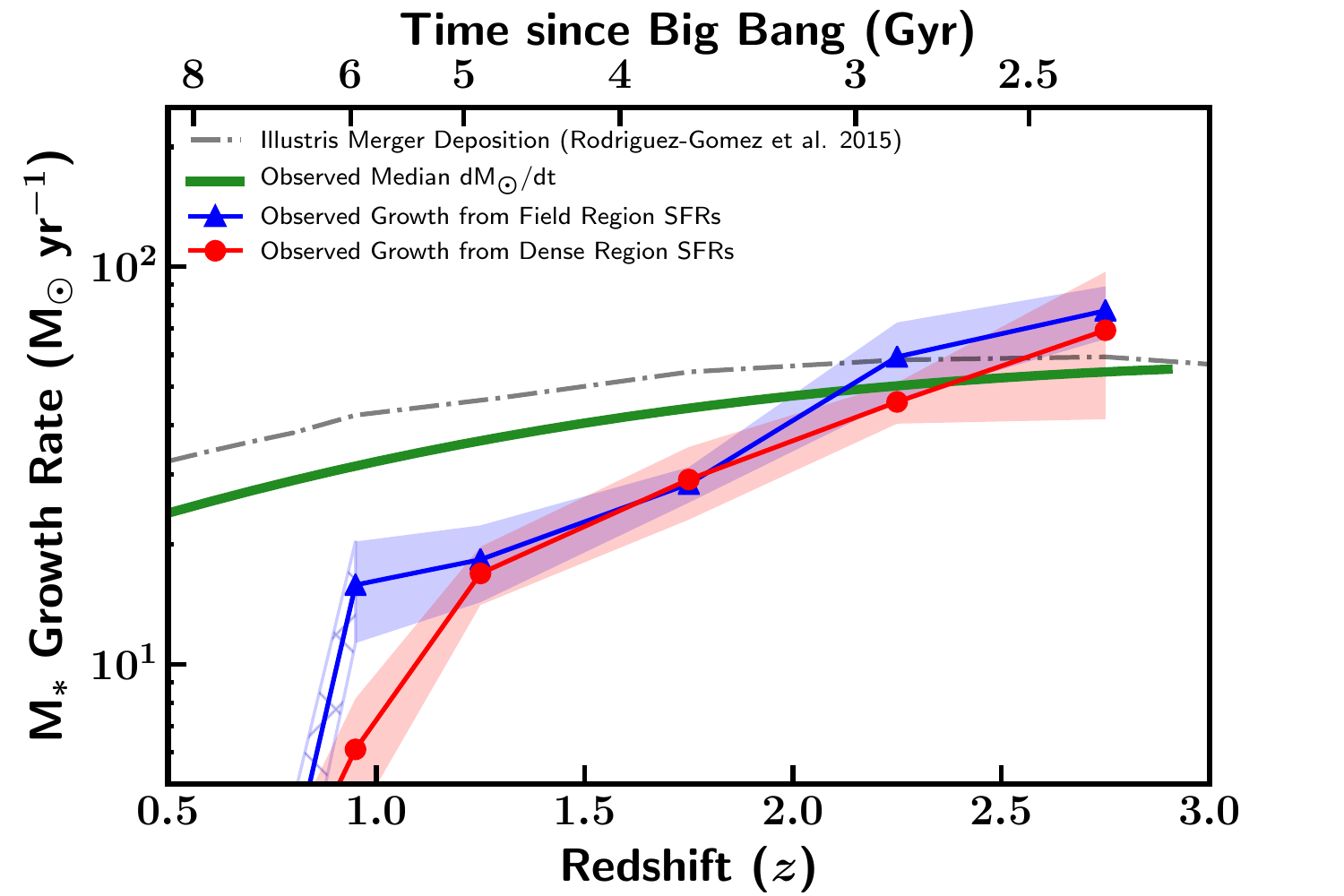}
\end{tabular}
\end{center}
\caption
{ \label{fig:illustrisgrowth} Observed stellar mass growth rate for massive galaxy progenitors in dense regions (red) and field regions (blue) compared to the median growth rate of our sample's median stellar masses (green). The galaxies selected in dense environments are those most likely to form the BCGs we see today, while those in the field sample are candidates to merge into a BCG during cluster mergers at later times.  We overplot the stellar mass deposition rate of mergers in the Illustris simulation (gray dashed--dotted) with a stellar mass capture efficiency of 100$\%$.  The hashed blue region in our lowest-redshift bin ($0.5 < z < 0.8$) indicates that only one progenitor was in a field environment.  We identify three epochs of stellar mass growth; an SF dominated phase at $z > 2.25$, a phase between $1.25 < z < 2.25$ where SF is responsible for $\sim$50\% of the total stellar mass growth and gas-rich and poor mergers are required to match the total (green), and finally a dry merger-dominated phase at $z < 1.25$ where star formation is insignificant to the total stellar mass growth.}
\end{figure*}

Our goal is to investigate what role in situ SF plays in the total growth of stellar mass in BCG progenitors.  To do so, we need to compare the growth solely due to SF to a measurement of total stellar mass growth.  We define our total growth rate to be the change in median stellar mass between each redshift bin in units of $M_{\odot}$ yr$^{-1}$.   We then perform a polynomial fit to these growth rates at each bin boundary and plot the result as the solid green line in Figure~\ref{fig:illustrisgrowth}.  

We determine the growth rate of each progenitor due to in situ SF by taking the SFRs from our SED fits and applying a reduction of 50$\%$.  This scale factor used by \citet{Brinchmann:2004aa} and \citet{van-Dokkum:2010aa} corrects for stellar mass loss over a Gyr timescale.  This correction to SFR produces a stellar mass growth measurement that considers the net growth influenced by SF and star destruction processes.  Stellar mass loss occurs as the population of stars born during an SF episode lose their massive members over time as they age off the main sequence.  The final in situ growth rate represents the total stellar mass per year generated in a galaxy that will remain on multi-Gyr timescales.   After we apply this correction, we calculate the in situ growth rate for progenitors in dense \big( $\frac{\delta}{\delta_{\mathrm{median}}}  >$ 2 \big) and field environments \big($\frac{\delta}{\delta_{median}} <$ 1 \big) environments and plot the mean stellar mass growth rates due to star formation for each sample as red and blue lines, respectively, in Figure~\ref{fig:illustrisgrowth}. $\delta$ is defined as the density at that location while $\delta_{\mathrm{median}}$ is the median density of the redshift slice.

Finally, we wish to understand how the stellar mass directly delivered via gas-rich and gas-poor mergers compares to the above values.  Therefore we integrate the stellar mass deposition rate ($M_{\odot}$ yr$^{-1}$) from mergers of secondary galaxies with mass ratios of 1:1 to 1:10$^{-5}$ of the progenitor's stellar mass in that redshift bin using the merger rates provided by the Illustris simulation \citep{Rodriguez-Gomez:2017aa}.  This is plotted as the dashed gray line in Figure~\ref{fig:illustrisgrowth}.

Following \cite{Hill:2017aa}, we plot the above values as $dM/dt$ ($M_{\odot}$ yr$^{-1}$) versus redshift from SF, mergers,  and all sources in Figure~\ref{fig:illustrisgrowth}. From this comparison, we identify three epochs of BCG progenitor growth.

\subsubsection{The SF-Dominated Epoch ($z > \mathit{2.25}$)}\label{sec:era1}
At $z > 2.25$, the in situ SF estimated in progenitors in all environments is consistent with the total growth rate of the median stellar mass of our sample.  This consistency indicates that progenitor growth is dominated by active SF with contributions of direct stellar mass injection and gas via gas-rich and gas-poor mergers.  The total stellar mass growth rate shown in Figure~\ref{fig:illustrisgrowth} is motivated by the stellar mass selection function that was used to originally select our sample; however, we may still conclude that the active SFRs via secular or wet merging processes are high enough to be the primary method of individual progenitor growth.  Dry mergers are unnecessary, but may still contribute to the highest-mass cases.

\subsubsection{The Transitionary Epoch ($\mathit{1.25} < z < \mathit{2.25}$)}\label{sec:era2}
Between $1.25 < z < 2.25$, progenitors are forming their in situ stellar mass at a rate $\sim$60--75$\%$ of the observed median total mass growth rate.  In this era, progenitors are still forming their mass predominantly through in situ SF independent of environment; however, additional methods of stellar mass generation are required.  It is during this time that the total stellar mass delivered by mergers becomes important to BCG progenitor evolution, as the mass formed via SF (red/blue lines in Figure~\ref{fig:illustrisgrowth}) is insufficient to account for the observed total stellar mass growth rate (green line in Figure~\ref{fig:illustrisgrowth}).

\subsubsection{The Dry Merger Epoch ($z < \mathit{1.25}$)}\label{sec:era3}
Finally, at $z < 1.25$ the in situ SF of BCG progenitors in dense environments sharply declines with redshift, while massive field galaxies maintain SF activity down to lower-redshifts.  In our lowest redshift bin ($z = 0.55$), the limited area of the COSMOS field introduces small-number statistics.  In this bin, we observe only one galaxy above our mass cut in a field environment, compared to 16 in dense environments. We see a general trend where SF is more than an order of magnitude insufficient to account for the total stellar mass growth rate.  This era requires ex situ mass delivery systems, specifically dry mergers, to be the dominant growth mechanism.

\begin{figure*}[t]
\begin{center}
\begin{tabular}{c}
\includegraphics{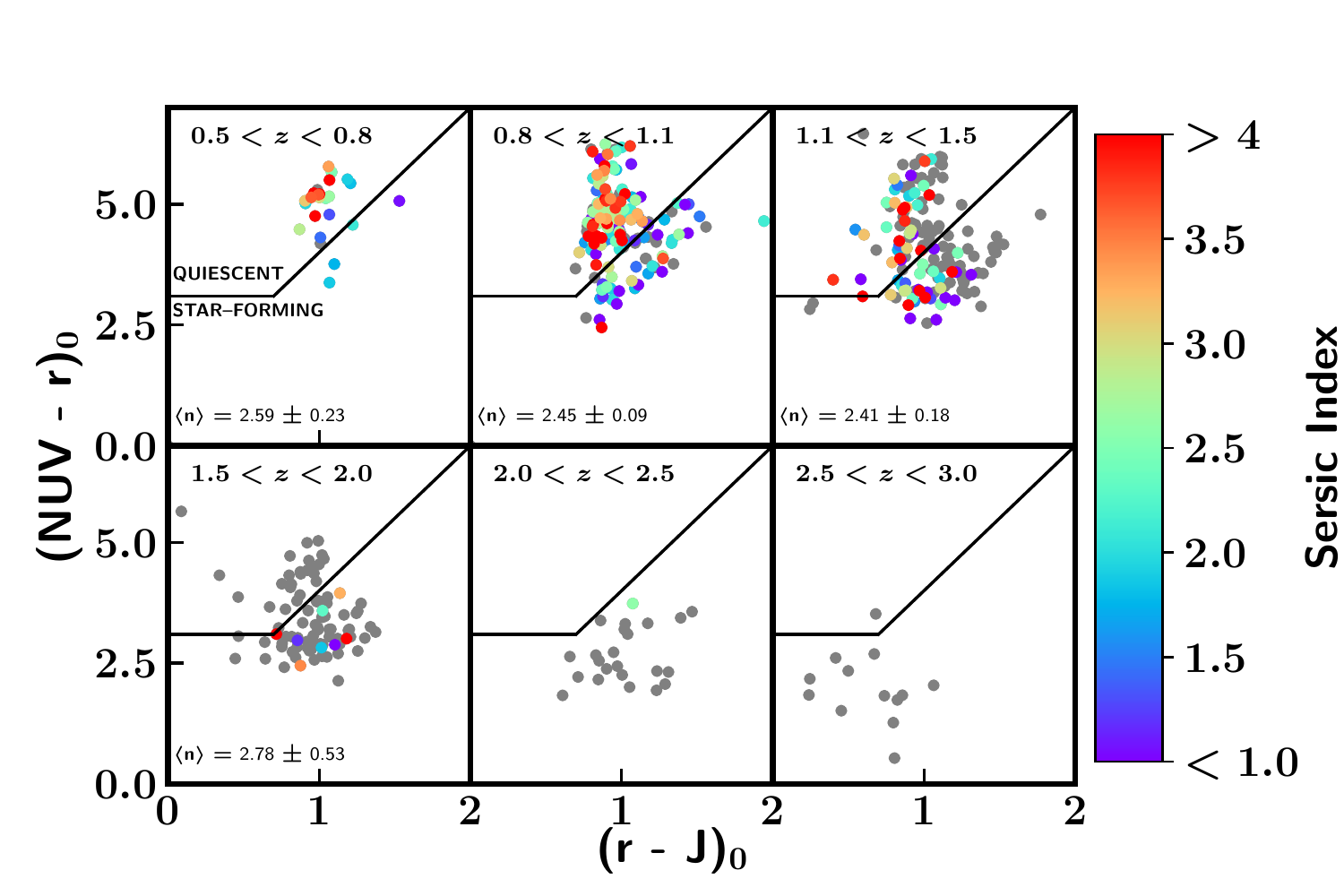}
\end{tabular}
\end{center}
\caption
{ \label{fig:nuvrj} Evolution of the Illustris-selected BCG progenitor sample in rest-frame NUV\emph{rJ} space.  Quiescent galaxies are selected by the top left quadrant indicated by the black lines.  Targets outside this region are considered star-forming.  Gray points are used for progenitors identified with a junk flag, used for spurious measurements or for targets too faint to fit a S\'ersic index \citep[$I_{AB}>$22.5;][]{Sargent:2007aa}. The median S\'ersic index of the overall population (bottom left of each panel) is constant within errors across all redshift bins with more than one measured index.   We find that the majority of progenitors selected by Illustris are star forming until $z \sim 1.5$, after which our progenitor sample is predominantly resides in a quiescent colorspace.}
\end{figure*}

\subsection{The Onset of Quenching}\label{sec:quenching}
We contextualize the diminishment of in situ SF toward progenitor growth by examining how rest-frame colors and morphological parameters such as the S\'ersic index evolve with redshift.  Shown in Figure~\ref{fig:nuvrj}, we plot the rest-frame NUV-\emph{r} and \emph{r--J} colors \citep{Ilbert:2013aa} of our progenitor sample in each of our six redshift bins with S\'ersic index from the COSMOS Zurich Morphological Catalog \citep{Sargent:2007aa,Scarlata:2007aa} as the colorbar. We use the rest-frame absolute magnitudes provided by the COSMOS2015 catalog.  Rest-frame magnitudes were calculated using the observed nearest filter magnitude and a $k$-correction estimated from the SED fitting performed by \citet{Laigle:2016aa}. Our SFR estimates are consistent with their target's NUV\emph{rJ} classification, with the star-forming galaxy subpopulation hosting median SFRs 10 times the SFR of the quiescent population in our lowest-redshift bin, and five times the SFR of the quiescent population in our highest-redshift bin.  A caveat remains that this diagnostic may misclassify star-forming galaxies as quiescent, and contaminate $>10\%$ of quiescent selected galaxies \citep{Ilbert:2013aa}.  Therefore we compare the total number of galaxies classified as quiescent to the number of quiescent galaxies above the star-forming main sequence, the correlation between SF and stellar mass indicating the median star-forming activity for a galaxy of a given stellar mass and redshift \citep[e.g.,][]{Schreiber:2015aa}.  We find that $\sim$3-5\% of progenitors classified as quiescent lie above the star-forming main sequence, potentially the misclassified star-forming galaxies found in \citet{Ilbert:2013aa}.  This contamination rate is low enough that the results shown in Figure~\ref{fig:nuvrj} are not affected.

S\'ersic indices were estimated using the HST/ACS F814W observations of the COSMOS field. NUV\emph{rJ} is more sensitive to recent SF evolution than UVJ-derived color measurements \citep{Martin:2007aa,Davidzon:2017aa}, an important consideration in determining when a galaxy has begun quenching. Position in this color space has also been shown to correlate with sSFR as the galaxy's star forming efficiency changes \citep{Martin:2007aa,Arnouts:2013aa,Ilbert:2015aa}.  We find that the stellar populations of our BCG progenitors transition to a quiescent state between our $0.8 < z < 1.1$ and $1.1 < z < 1.5$ redshift bins, including those without morphological classifications.  This is consistent with our results from Figure~\ref{fig:illustrisgrowth}, where in situ SF becomes a negligible contributor to a BCG's evolution between $1.0 < z < 1.3$.  Cotemporal with the transition to a quiescent stellar population, the morphology of the progenitor population also changes.  The median S\'ersic index in each redshift bin (Figure~\ref{fig:nuvrj}) is constant within the scatter, centered at $\langle$n$\rangle \sim 2.5$.  These median S\'ersic indices are consistent with those found for S0 galaxies in cluster environments \citep[e.g.,][]{D_Onofrio_2015}, suggesting that galaxies in our progenitor sample have hosted a composite bulge+disk structure since $z \sim 2$.

\subsection{Visual Classification of Morphology}\label{sec:diskiness}
To provide context to the light profile evolution observed in Section~\ref{sec:quenching}, we classify the morphological state of all progenitors imaged in the COSMOS field ACS mosaic \citep{Koekemoer:2007aa}, totaling 1372 of our original sample of 1444.  Progenitors not included are outside the bounds of the ACS mosaic.  The COSMOS ACS mosaic includes HST/ACS F814W broadband filter observations with a final pixel scale of 0.03\arcsec\;per pixel and a 3$\sigma$ surface brightness density of $m_{AB}$ $\sim$ 25.1 mag arcsec$^{-2}$ \citep{Wen-wenzhangzheng:2016aa}.  This spatial resolution corresponds to 183 pc per pixel at $z \sim 0.5$ and 231 pc per pixel at $z \sim 3$.  We also supplement this observation set with any publicly available images from HST/WFC3 F105W, F110W, F125W, F140W, and F160W.   This subset of 381 targets is most useful in our high-redshift bins as the peak of the stellar continuum is redshifted into the WFC3 IR broad bands.  The following results shown in this work are consistent independent of the inclusion or exclusion of targets with WFC3 observations.  To ensure consistency of classification, we utilize the morphological classifications scheme and user interface implemented by the CANDELS team \citep{Kartaltepe:2015aa,Kocevski:2015aa} to classify each galaxy's morphology (spheroid, disk, irregular) as well as interaction class (merger, interaction, non-interacting companion).  

\begin{figure}
\begin{center}
\begin{tabular}{c}
\includegraphics[width=0.51\textwidth]{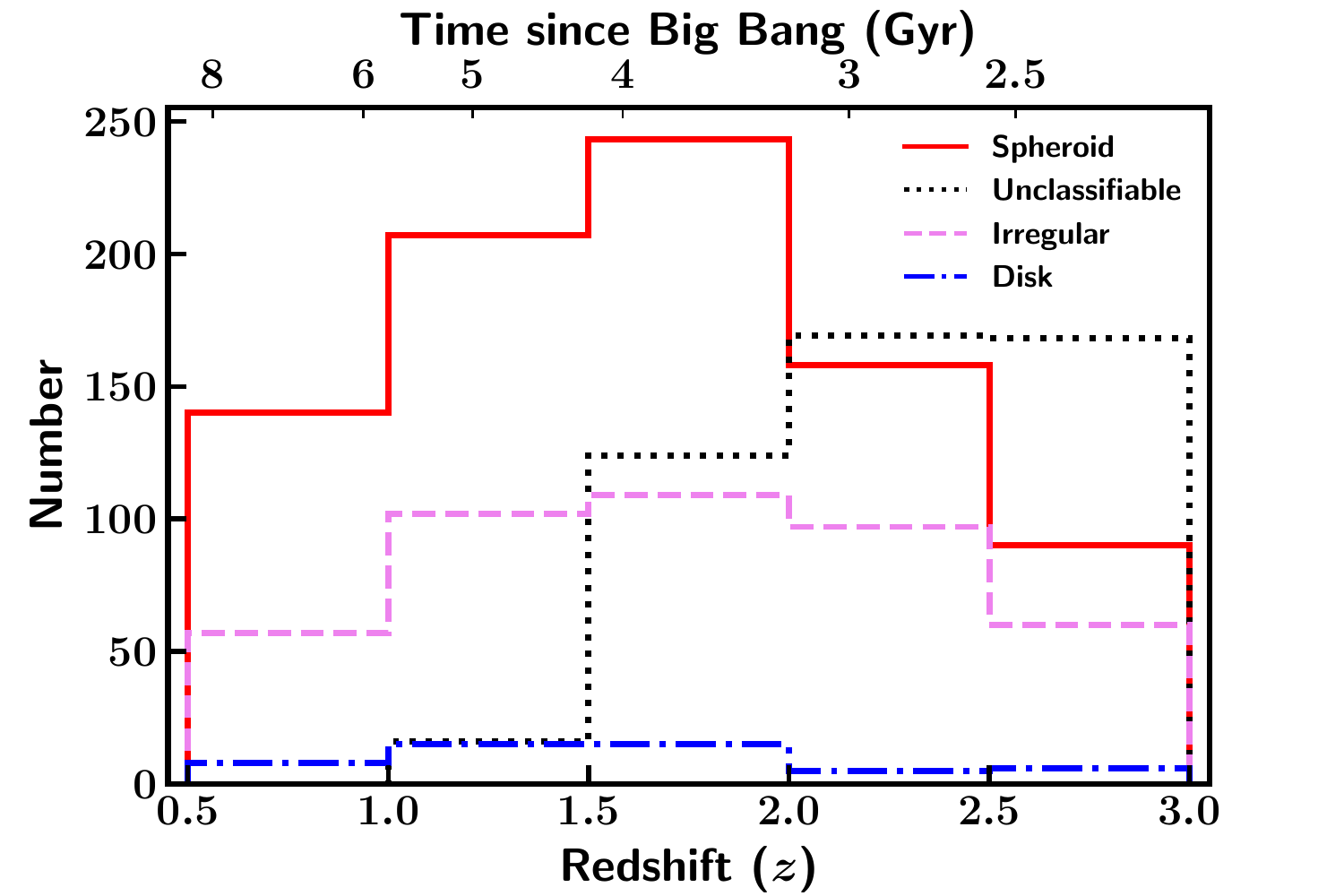}
\end{tabular}
\end{center}
\caption
{ \label{fig:morphtypehist} Distribution of morphological types (spheroid, disk, irregular, unclassifiable) for our sample of BCG progenitors.  To be classified as spheroid, disk, irregular, or unclassifiable two out of three classifiers must agree on the designation and each subsample may include members of others if two labels are used by two classifiers.}
\end{figure}

The user examines the HST image and labels the progenitor with as many flags as necessary to classify it. Our sample was classified by three authors (K.C.C., J.S.K., K.D.T.), and two of three authors must agree on a designation to assign a galaxy to any of the classification bins.  For galaxies that lack the organized, symmetric morphology of a disk and/or spheroid, the irregular flag is used.  To provide context to the primary classification, interaction classes are also used for the apparent stage of an interaction. The merger flag is used in highly disturbed cases where the primary and secondary galaxies of the mergers are nearly indistinguishable.  The interaction flag is used for disturbances such as tidal tails where the interacting galaxies remain independently resolved.  For more details on this classification scheme see \citet{Kartaltepe:2015aa}.  The distribution of morphologies for our BCG progenitor sample is shown in Figure~\ref{fig:morphtypehist}. Our sample is predominantly spheroidal at low redshift, with a growing fraction of unclassifiable progenitors toward high redshift, where the rest-frame optical stellar continuum is redshifted out of the observed band.  We see the majority of progenitors are spheroidal, indicating that a significant fraction of our total progenitor sample has already developed the spheroidal structure exhibited by present day BCGs.  This appears to be consistent with recent work that identifies a population of massive spheroidals at high redshift \citep[e.g.,][]{Cassata:2013aa}.

\begin{figure}
\begin{center}
\begin{tabular}{c}
\includegraphics[width=0.51\textwidth]{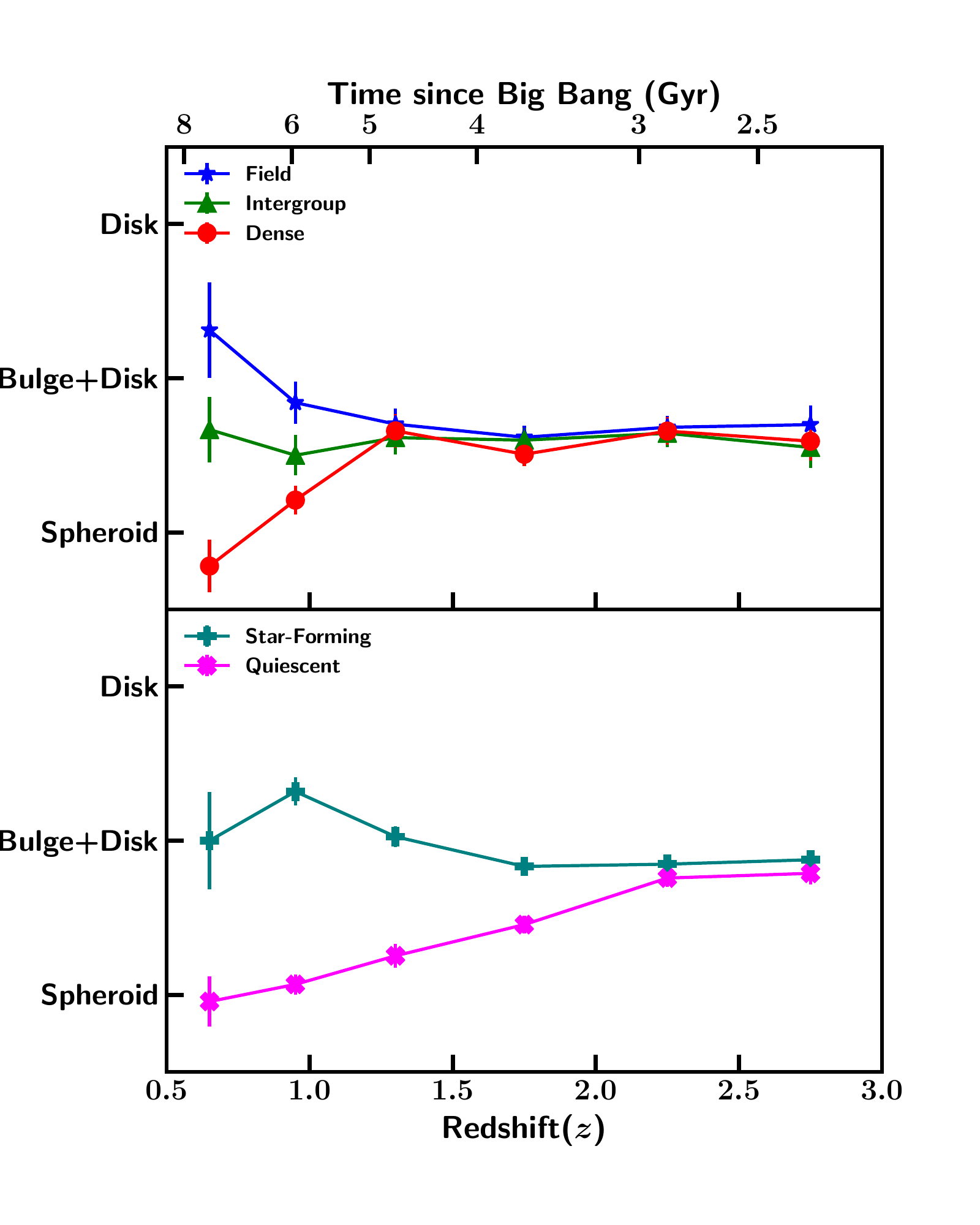}
\end{tabular}
\end{center}
\caption
{ \label{fig:morphenvshist} Top: median diskiness parameter, described in Section~\ref{sec:diskiness},  for our progenitor sample binned by field, intergroup, and dense environments.  We identify a morphological dependence on environment at low redshift.   However, it is difficult to classify fainter, more dust-obscured galaxies at high redshift. Bottom: median diskiness parameter with redshift for progenitors identified as quiescent (magenta) and star-forming (teal) using their NUV\emph{rJ} colors.  We find the quiescent population evolves to a spheroid-dominated population with time, while star forming galaxies remain a combination of spheroid and disk. We caution that the ACS F814W images predominantly used here will be dominated by the rest-frame UV past $z > 1$.}
\end{figure}

To compare the environmental effects discussed in Sections~\ref{sec:era1}--\ref{sec:era3}, we plot the net classification, or `diskiness' of progenitors in our three environment bins (top half of Figure~\ref{fig:morphenvshist}). Our diskiness parameter is defined by the averaging of disk flags (valued as +1) and spheroid flags (valued as -1).  For example, a target with two spheroid flags and one disk flag will have a diskiness value of -0.33. Flags indicating disk- or bulge-dominated features are also included as values of $\pm$0.5, respectively. These flags are used to refine the classification of a galaxy with multiple primary flags and identify objects such as bulge-dominated S0 galaxies.  We find that low-redshift progenitors have a decreasing likelihood of being labeled a disk with increasing local density. This trend is muddled in our highest-redshift bins, where we do not observe evidence for an environmental dependence.   Next, we plot the diskiness parameters of progenitors identified as star-forming and quiescent through their NUV\emph{rJ} colors (see Section~\ref{sec:quenching}) in the bottom half of Figure~\ref{fig:morphenvshist}. We find a clear offset of average net classification for these two subsamples, where quiescent galaxies are more spheroidal on average than their star-forming counterparts below $z \sim 2.25$. 

We next consider how often progenitors are classified as an irregular with redshift, and how this flag is used in combination with other flags such as disk or spheroid.  We include example HST/ACS F814W images of the possible irregular combinations in Figure~\ref{figure:examplemorphs}. In Figure~\ref{fig:crossmorph}, we plot the fraction of the total sample that has at least two irregular flags per target, identical to the confident irregular subsample in Figure~\ref{fig:morphenvshist}.  This total irregular sample is divided into six redshift bins with subsamples that include at least one spheroid, disk, or composite (spheroid+disk) flag.  As these are confident irregulars (two or more irregular flags), we lower the threshold of additional flags to one to examine how often other characteristics such as disks or spheroids are assigned to this population.  The frequency of irregulars in our sample declines with decreasing redshift.  Also with redshift, we see the irregular population transition from a diverse range of morphologies to an irregular-spheroid and irregular-composite population toward present day.  One result of interest is the lowest-redshift bin, in which disk components are identified in all the irregular candidates, a rare feature ($\leq$ 5\%) in the total sample at $z < 1$.

\begin{figure*}
\epsfig{file=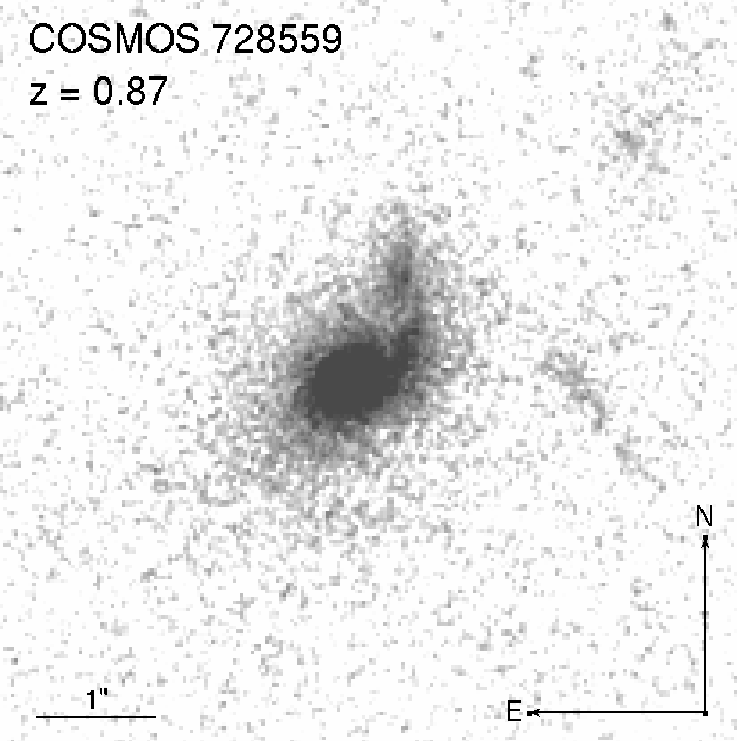,width=0.49\textwidth,angle=0}
\epsfig{file=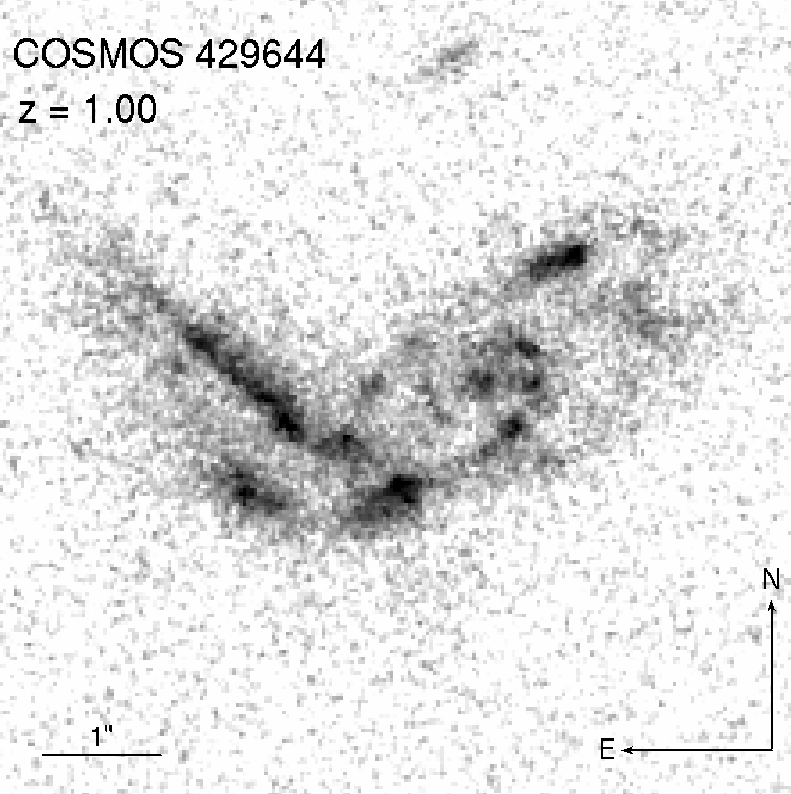,width=0.49\textwidth,angle=0}
\end{figure*}
\begin{figure*}
\epsfig{file=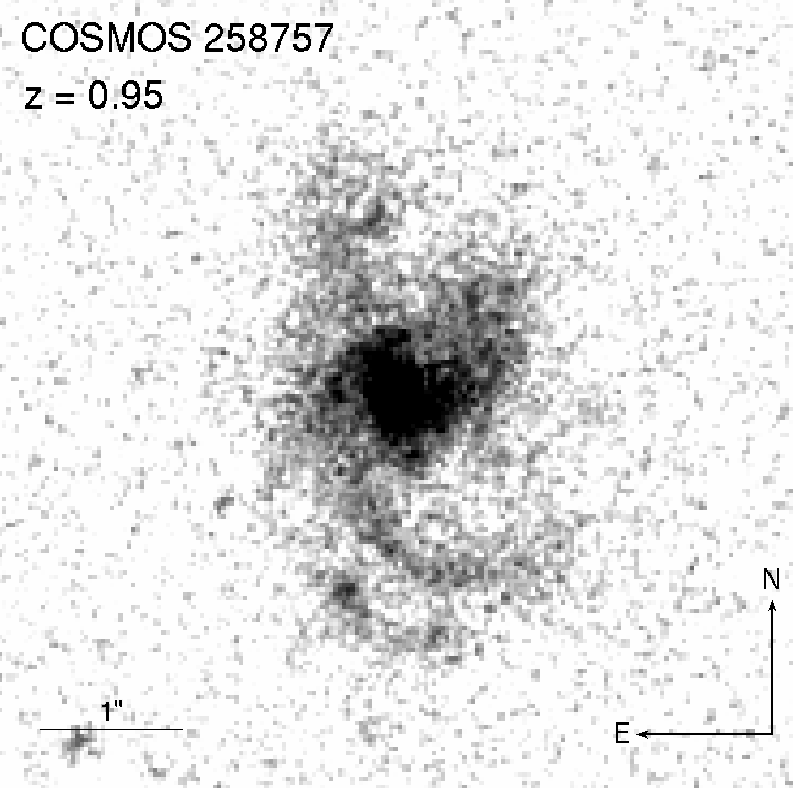,width=0.49\textwidth,angle=0}
\epsfig{file=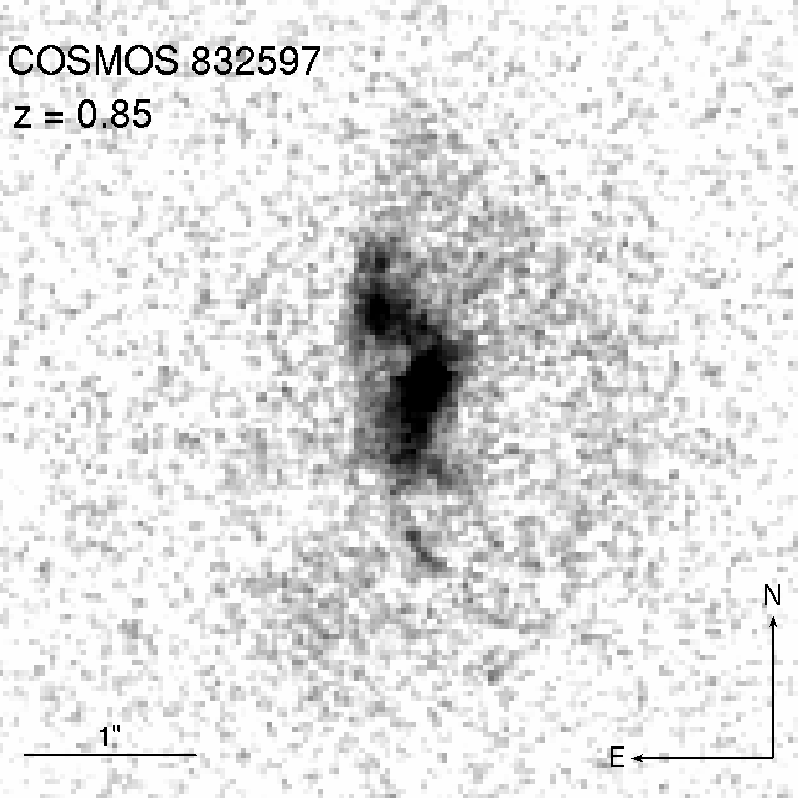,width=0.49\textwidth,angle=0}

\caption{\label{figure:examplemorphs} 
Example irregular galaxies from the HST/ACS F814W broadband filter mosaic of the COSMOS field, identified by their ID in the COSMOS2015 catalog and photometric redshift \citep{Laigle:2016aa}.  Top left:  An irregular spheroid with a tail to the northwest of the nucleus. Top right: an irregular disk with a predominantly disk-like structure and a tidal tail to the west of the disk, potentially indicating an ongoing merger. Bottom left:  A composite disk+spheroid irregular, with two spiral arms to the south of a spheroid component.  Bottom right:  a purely irregular galaxy with an asymmetric appearance devoid of spheroidal core or spiral arms. }
\end{figure*}

\begin{figure}
\begin{center}
\begin{tabular}{c}
\includegraphics[width=0.51\textwidth]{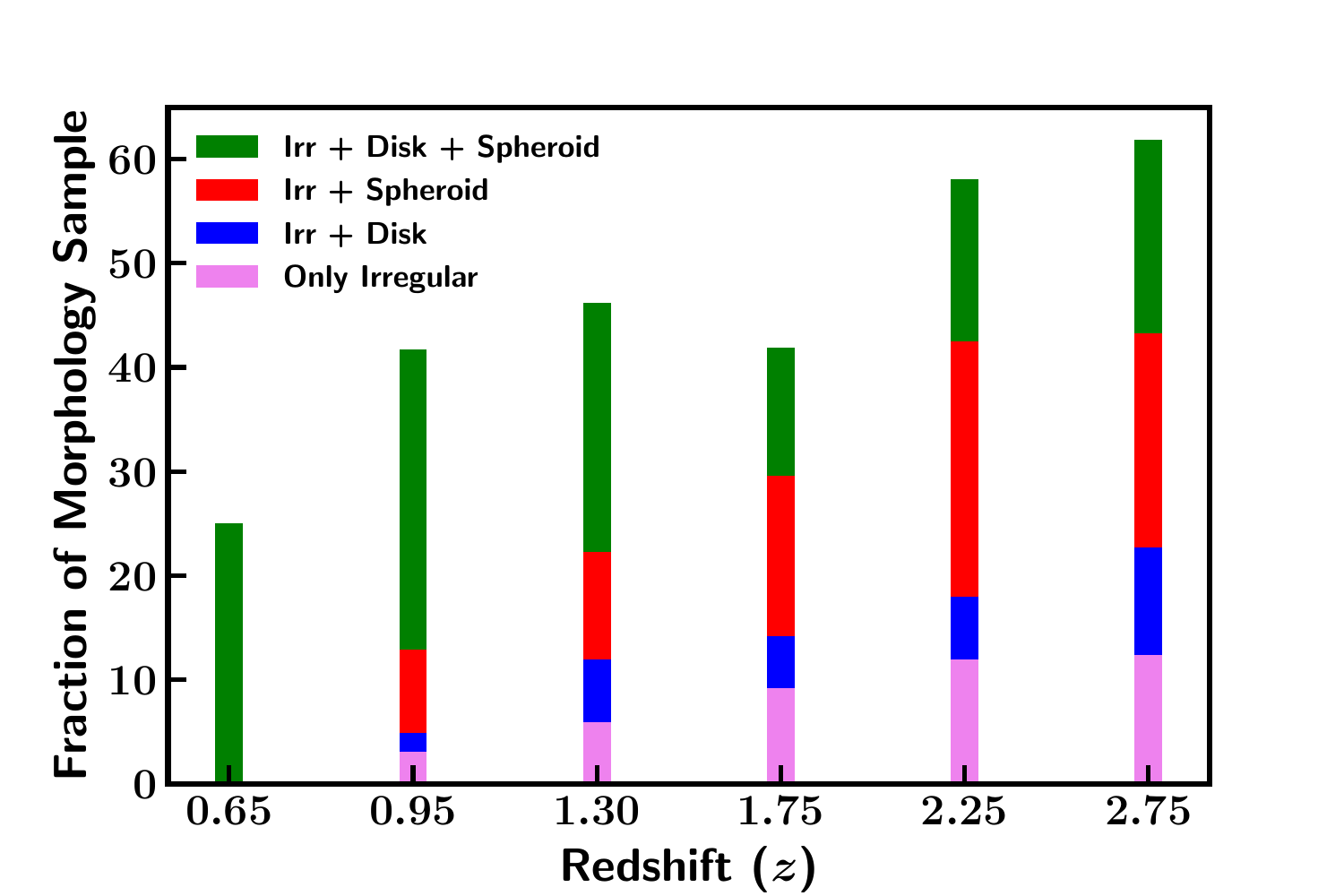}
\end{tabular}
\end{center}
\caption
{ \label{fig:crossmorph} Fraction of our visually classified progenitor sample which were labeled as irregulars, using ``only irregular" (pink), ``irregular+disk" (blue), ``irregular+spheroid" (red), or all three component (green) flags.}
\end{figure}

\begin{figure}
\begin{center}
\begin{tabular}{c}
\includegraphics[width=0.51\textwidth]{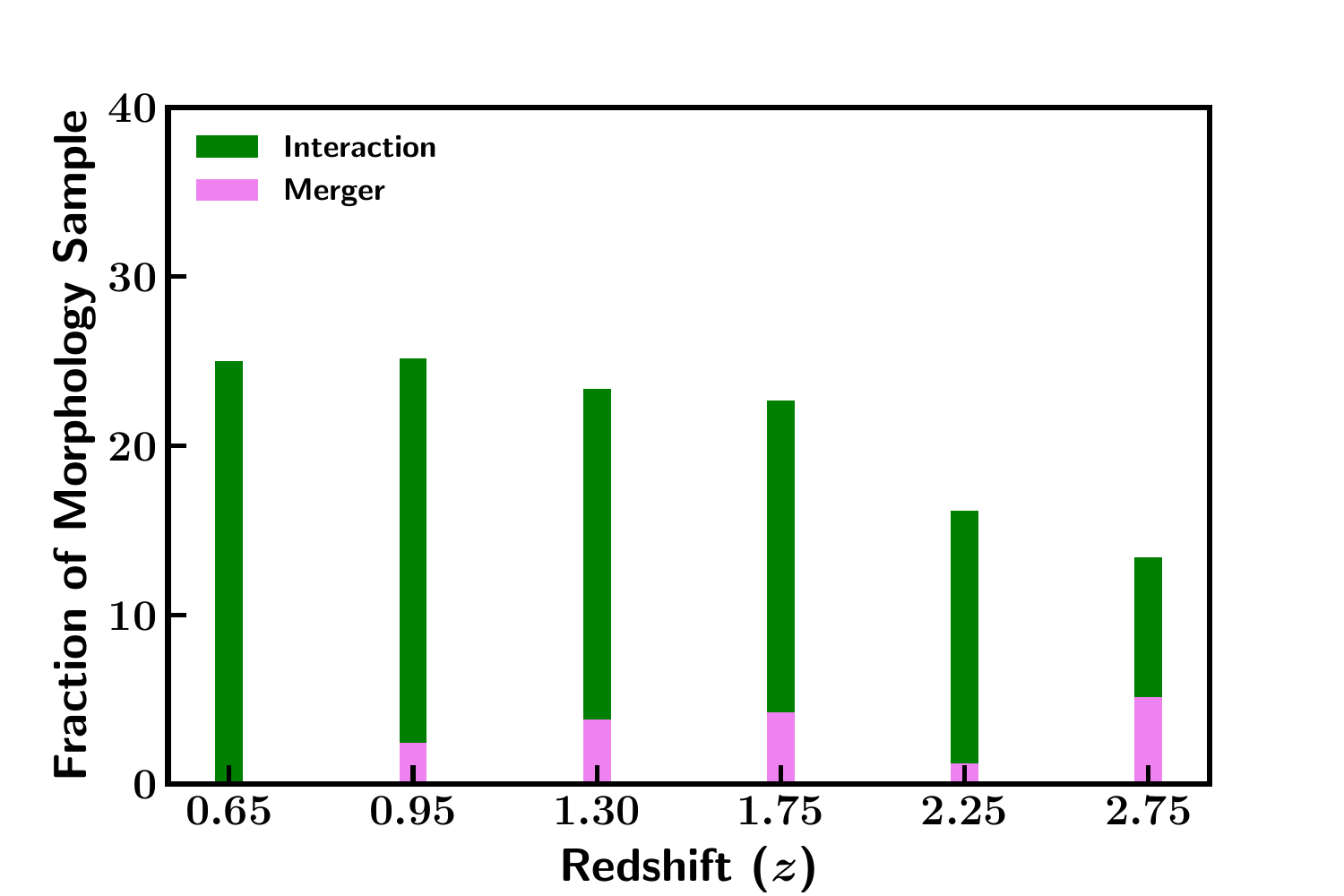}
\end{tabular}
\end{center}
\caption
{ \label{fig:mergersonlymorph} Fraction of our visually classified progenitor sample which included any interaction flags, using ``Interaction" (green), and ``Merger" (pink) flags.  Mergers are differentiated from interactions by the degree to which the observer can distinguish the two interacting galaxies from each other.}
\end{figure}

Finally, we examine the distribution of interaction classifications (Figure~\ref{fig:mergersonlymorph}).  To classify the state of interactions, we use flags to identify any ongoing `mergers' in which the structure of both galaxies have been disturbed and combined.  An `interaction' flag is used when an interacting pair, identified using tidal features, are visibly distinct from one another.  We note that these classifications do not distinguish between major and minor merger pairs. We observe a slowly decreasing interaction fraction with redshift ($\sim 25-15\%$), with a roughly constant merger fraction of $5\%$ of the total sample.  We find a $1\sigma$ Poisson confidence interval of $\pm 5\%$ between $0.5 <  z < 2.0$ and  $\pm 3\%$ between $2.0 < z < 3.0$. Our results are consistent within errors of those calculated by \citet{Man:2016aa} for the general galaxy population; they find a total interaction frequency of $\sim$20\% at $z \sim 1$ using a lower-mass sample of pairs down to $10^{10.8}$ $M_{\odot}$. The lack of observed mergers in our lowest-redshift bin is consistent with a low-redshift, minor merger-dominated model for BCG growth \citep[e.g.,][]{Edwards:2012aa}.  It is also consistent with the results from \citet{Duncan:2019aa}, who found a merger fraction of $<10\%$ for their massive subsample (10$^{10.3}$ $M_{\odot}$) in the CANDELS fields.  The lack of interactions in our highest-redshift bin may be due to the difficulty in visually identifying low surface brightness features present in interactions at $z > 2$.  Overall this slowly evolving fraction of interactions is consistent with a model with fewer mergers at later times.

\subsection{Comparison to Literature}
Several works on massive galaxy evolution and the comparison of progenitor selection methods have been published in recent years \citep[e.g.,][]{Zhao:2016aa,Hill:2017aa, Torrey:2017aa}, with results consistent with ours. \citet{Hill:2017aa} selected massive galaxy progenitors using the number density evolution functions derived from the Bolshoi simulation \citep{Klypin:2011aa}.  Due to the dark-matter-only nature of the Bolshoi simulation, a comparison between our work and that done by \citet{Hill:2017aa} helps illustrate how stellar mass selection functions derived using different combinations of dark and baryonic physics affect measured evolution.  

\citet{Hill:2017aa} used a lower stellar mass progenitor selection than ours.  Since lower-mass galaxies build their stellar mass over longer timescale on average \citep[mass downsizing;][]{Cowie:1996aa,Perez-Gonzalez:2008aa}, \citet{Hill:2017aa}'s selection led to an in situ SF-dominated phase which persisted to lower redshift than ours, until $z \sim 1.75$.  However, they found a fast transition to a quenched state from $z < 1.75$ \citep[Figure 1 of][]{Hill:2017aa}.   Dark-matter-only simulations producing a lower-mass selection function is hypothesized to be due to the lack of baryonic physics feedback effects that limit the speed of stellar mass growth, enabling the inclusion of faster evolving, lower-mass progenitors.

The median stellar mass of our sample increases by a factor of $\sim 80\%$ over $0.35 < z < 2.00$, and a factor of two when extrapolated to the present day from $z = 2$.  This rate of BCG stellar mass growth is slower than the factors of two from $z = 1$ shown in BCGs residing in clusters selected from \emph{Spitzer} observations \citep{Lidman:2012aa,Lin:2013aa,Shankar:2015aa}.  However, our results are in rough agreement with \citet{Zhao:2016aa}, who found that BCG progenitors grow by a factor of 2 from $z \sim 2$ to present day.  The \citet{Zhao:2016aa} sample was selected using a hybrid selection function which selected the most massive galaxies in the 38 most overdense regions identified in the CANDELS UDS field. The \citet{Zhao:2016aa} sample also exhibited dry merger-dominated growth at $z < 1$.  Moreover, our work is consistent with additional works that indicate a slow rate of growth at low redshift \citep[e.g.,][]{Tonini:2012aa,Hill:2017aa}, shown in BCGs selected via the \citet{Maraston:2005aa} stellar population models or comoving number density cuts.  Our observation of a transition to a dry merger-dominated epoch below $z \sim 1$ is also consistent with the Spitzer Wide-Area Infrared Extragalactic (SWIRE) Survey \citep{Lonsdale:2003aa} samples of \citet{Webb:2015aa}.  However, above $z \sim 1$, our sample includes progenitors with SFRs an order of magnitude lower than those of \citet{Webb:2015aa}.  This is believed to be primarily due to the difference in depth of the 24$\mu$m imaging in COSMOS versus SWIRE (71 $\mu$Jy versus $\sim$150 $\mu$Jy, respectively).

\subsection{Comparison to the Millennium-II Simulation}\label{sec:millcompar}
As there is not yet a perfect cosmological simulation of the universe, it is worthwhile to examine how choosing a different simulation may change our results.  Dark-matter-only simulations such as MS-II and Bolshoi, for example, have a different merger tree history than Illustris, which changes the rate ex situ stellar mass is accumulated.  To test this, we retrieve the median cumulative number density of progenitor halos from MS-II \citep{Boylan-Kolchin:2009aa}.   To identify BCG progenitors simulated in MS-II, we identify the halo population inhabiting a comoving volume at the same density as observed low-redshift BCGs \citep{Davidzon:2017aa}.   We then identify the most massive progenitor halo corresponding to each descendant halo in ascending redshift slices to $z \sim 3$ and measure the cumulative comoving number density corresponding to the median most massive halo in that redshift slice.  This number density is used to select mock galaxies from \citet{Guo:2013aa}, which represent the baryonic content of the MS-II BCG progenitor halos.  The median \citet{Guo:2013aa} sample stellar mass is used as to select observed galaxies in COSMOS based on their stellar mass (teal line in Figure~\ref{fig:evolutiontracker}).

\begin{figure}
\begin{center}
\begin{tabular}{c}
\includegraphics[width=0.51\textwidth]{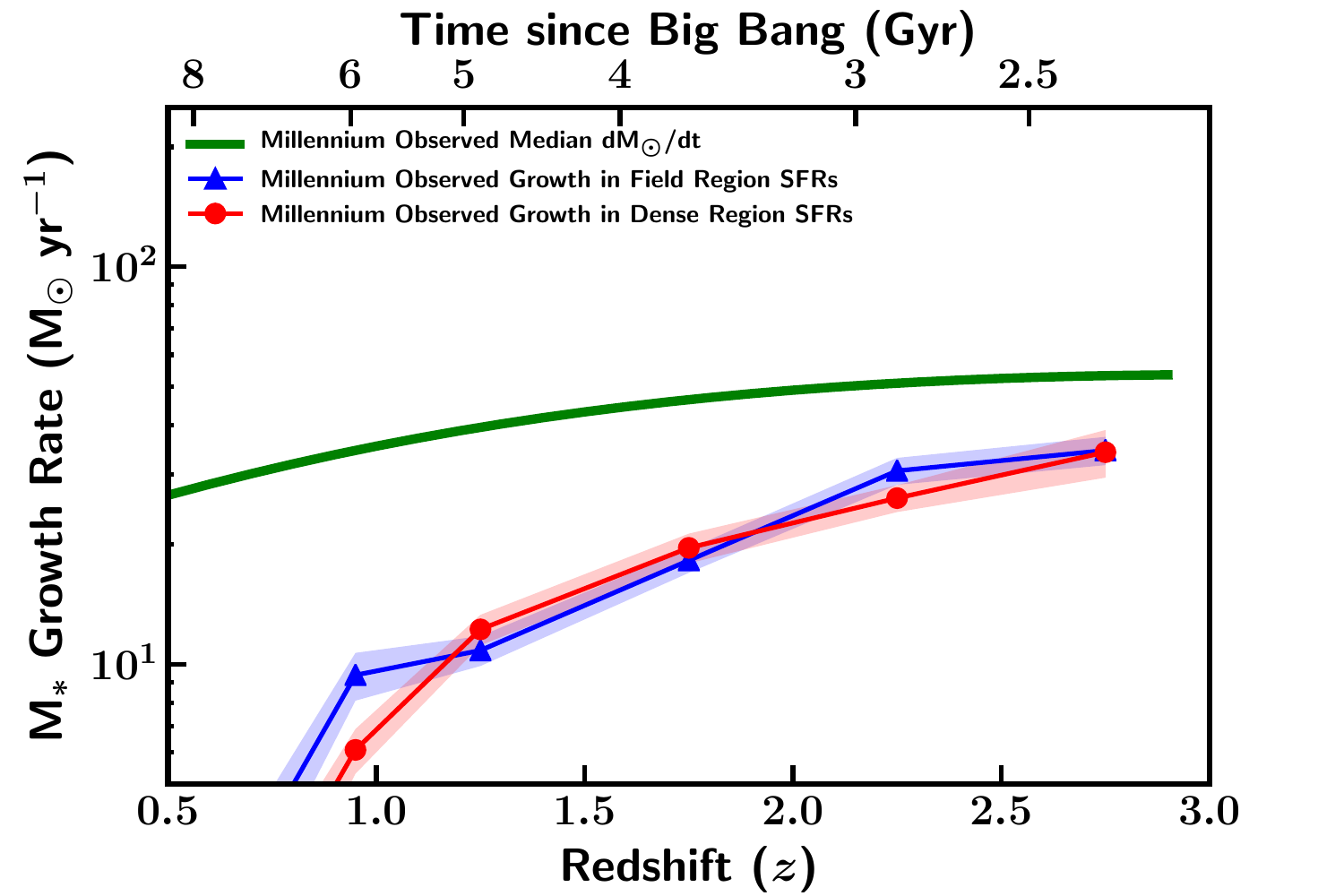}
\end{tabular}
\end{center}
\caption
{ \label{fig:millgrowth} Observed stellar mass growth rate for MS-II-selected massive galaxy progenitors in dense regions (BCG progenitors; red) and field regions (blue), and the median growth rate of our sample's stellar mass (green).}
\end{figure}

We fit the SEDs of MS-II-selected progenitors identified in the COSMOS2015 catalog in the same manner as the Illustris-selected progenitors, as described in Section~\ref{sec:methods}.  Shown in Figure~\ref{fig:millgrowth}, the in situ growth of massive galaxy progenitors due to star formation is at least half of the total growth until $z \sim 1.5$, lower than our findings using Illustris-selected progenitors.  However, the total growth rate track has a much higher normalization than Illustris, indicating a later, faster growth history in MS-II.  To be physically consistent between SFR and total growth in MS-II selected progenitors, a much richer merger history is required to deliver mass onto the simulated BCG seeds in addition to their intrinsic SF.  These differences between Illustris and MS-II may have several causes.  One is the difference in feedback physics between the MS-II \citep{Guo:2013aa} semi-analytical models assigned to their halos and the baryonic physics in Illustris.  A less efficient feedback process in MS-II would prevent the slow- or shut-down of highly star forming progenitors. This faster evolution produces a mass selection function with a larger slope, including galaxies with correspondingly lower SFRs.  

\subsection{Comparison to the Constant Density Method}\label{sec:constantcompar}

\begin{figure}
\begin{center}
\begin{tabular}{c}
\includegraphics[width=0.51\textwidth]{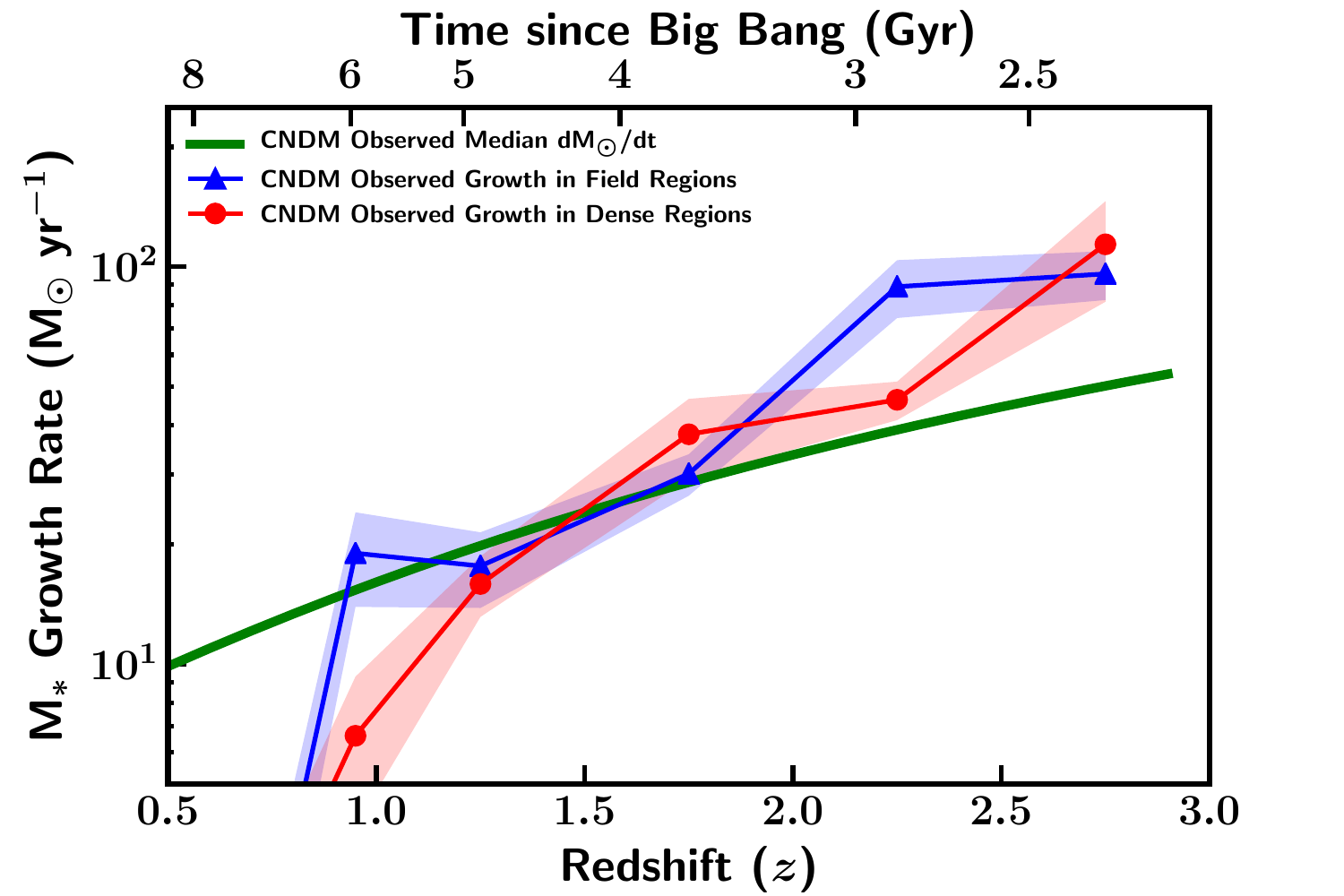}
\end{tabular}
\end{center}
\caption
{ \label{fig:constgrowth} Observed stellar mass growth rate for massive galaxy progenitors selected by constant number density in dense regions, i.e., BCG progenitors (red) and field regions (blue), and the median growth rate of our sample's stellar mass (green).}
\end{figure}

We also compare our results to the constant cumulative number density selection method \citep{van-Dokkum:2010aa}.  Following the constant number density selection function plotted in Figure~\ref{fig:evolutiontracker}, only the most massive progenitors are selected.  As discussed in Section~\ref{sec:methods}, a constant number density selection only selects progenitors that inhabit the universe at an equal density, an assumption that the number of galaxies in the universe is constant.  As this ignores the effect of mergers, the median progenitor mass is not pulled down by the lower-mass secondaries.  We use the same fits described in Section~\ref{sec:methods} to estimate the average stellar mass accretion via in situ SF in comparison with the evolution of the median stellar mass of the progenitor sample in Figure~\ref{fig:constgrowth}. We find that the combination of slow predicted stellar mass evolution and selection of very massive galaxies that are likely to be star forming result in a model where SF is dominant until $z < 1$, after which gas poor mergers are necessary. 

\section{Summary and Conclusions}\label{sec:conc}
We examine the in situ SF and ex situ stellar deposition components of the total stellar mass growth of progenitors for the present day BCG population.  By using FUV--FIR photometric data in the COSMOS field, we fit the SEDs for several samples of BCG progenitors and determine their SFR, sSFR, and stellar mass. Once corrected for stellar mass loss, we identify multiple epochs of BCG evolution based on the prominence, or lack thereof, of in situ SF toward stellar growth.  Our estimates of in situ SF include episodes fueled by a progenitor's own gas supply as well as those triggered by gas-rich mergers.  Total stellar mass growth also includes the direct contribution of stellar mass by dry mergers. These behaviors are then compared to local environment to test for the presence of any dependence.  Our results are as follows.

\begin{itemize}
\item{BCG progenitor growth is dominated by in situ SF, either secular or merger driven, until approximately $z \sim 2.25$, after which galaxy mergers (gas rich and gas poor) contribute increasing fractions of ex situ stellar mass growth with time until $z \sim 1.25$.  After this time, dry mergers become the dominant growth mechanism.}
\item{Any dependence on local environment at $z > 1$ is indistinguishable within our errors.  The SFR--local density correlation observed today is observed in our $z \sim 0.8$ bin; however, volume effects limit our ability to measure this effect at lower redshifts.  We also find that spheroidal progenitors become more common in dense environments \big($\frac{\delta}{\delta_{median}}  >$ 2\big) than field environments \big($\frac{\delta}{\delta_{median}}  <$ 1\big), which most often host spheroid+disk composites.}
\item{The era of quenching, either through merger or AGN activity, is encountered by most progenitors by $z \sim 1.00-1.25$.  This era is characterized by a shut-down of in situ SF and the transition from a disk+spheroid morphology to an overall spheroidal type.}
\item{Across all redshifts, progenitors requiring an AGN-fitting component represent a small fraction ($\sim$ 1\%) of the total progenitor population. SED3FIT AGN are most often observed at $z > 1.5$, with 29/37 of our SED3FIT AGN sample observed at high redshift.}
\item{Our progenitor sample consists of spheroid+disk composites until $z \sim 1.3$, after which we identify an environmental dependence on morphology.  Progenitors in dense environments are labeled with a more spheroidal net classification as redshift decreases.  This is primarily driven by the change in diskiness of the quiescent population in dense regions.   Progenitors in intergroup regions \big($1 < \frac{\delta}{\delta_{median}}  <$ 2\big) retain their composite structure. The few galaxies in field environments that meet our stellar mass cut increase in diskiness. Mergers and interactions are no more common than the general population ($< 25\%$), with irregular galaxies exhibiting a greater range of morphological classifications with higher redshift.}
\end{itemize}

\acknowledgements{
\linespread{1}
We thank the referee for the constructive feedback that improved the clarity of this work. Support for this work was provided by NASA through grants \emph{HST}-GO-13657.010-A and \emph{HST}-AR-14298.004-A awarded by the Space Telescope Science Institute, which is operated by the Association of Universities for Research in Astronomy, Inc., under NASA contract NAS 5-26555. Support and observations made with the NASA/ESA \emph{Hubble Space Telescope}, obtained from the Data Archive at the Space Telescope Science Institute, which is operated by the Association of Universities for Research in Astronomy, Inc.,  were also provided by NASA through grant NNX16AB36G as part of the Astrophysics Data Analysis Program.  Spectral energy distribution fitting was performed using the scientific computing resources of the Rochester Institute of Technology ION cluster.  B.D. acknowledges financial support from NASA through the Astrophysics Data Analysis Program (ADAP), grant No. NNX12AE20G, and the National Science Foundation, grant No. 1716907.  We thank Paul Torrey for his council on evolving number density methodology. We thank the generous advice of Elisabete da Cunha and Stefano Berta on the use of MAGPHYS and SED3FIT, respectively. This research made use of Astropy,\footnote{http://www.astropy.org} a community-developed core Python package for Astronomy \citep{Robitaille:2013aa,Price-Whelan:2018aa}.  This research made use of the iPython environment \citep{Perez:2007aa} and the Python packages SciPy \citep{Jones:2001aa}, NumPy \citep{Walt:2011aa}, and Matplotlib \citep{Hunter:2007aa}.
 
 Based on observations made with the NASA \emph{Galaxy Evolution Explorer}. \emph{GALEX} is operated for NASA by the California Institute of Technology under NASA contract NAS5-98034.  MegaPrime/MegaCam, a joint project of CFHT and CEA/DAPNIA, at the Canada-France-Hawaii Telescope (CFHT) which is operated by the National Research Council (NRC) of Canada, the Institut National des Sciences de l'Univers of the Centre National de la Recherche Scientifique of France, and the University of Hawaii.  Based [in part] on data collected at Subaru Telescope, which is operated by the National Astronomical Observatory of Japan. Based on observations obtained with WIRCam, a joint project of CFHT, Taiwan, Korea, Canada, France. This work is based [in part] on observations made with the \emph{Spitzer Space Telescope}, which is operated by the Jet Propulsion Laboratory, California Institute of Technology under a contract with NASA.  \emph{Herschel} is an ESA space observatory with science instruments provided by European-led Principal Investigator consortia and with important participation from NASA. 
 
This work was supported by a NASA Keck PI Data Award,
administered by the NASA Exoplanet Science Institute. Data
presented herein were obtained at the W. M. Keck Observatory
from telescope time allocated to the National Aeronautics and
Space Administration through the agency's scientific partnership
with the California Institute of Technology and the University of
California. The Observatory was made possible by the generous
financial support of the W. M. Keck Foundation. The authors
wish to recognize and acknowledge the very significant cultural
role and reverence that the summit of Mauna Kea has always had within the indigenous Hawaiian community. We are most
fortunate to have the opportunity to conduct observations from
this mountain.

}

\newcommand\invisiblesection[1]{%
  \refstepcounter{section}%
  \addcontentsline{toc}{section}{\protect\numberline{\thesection}#1}%
  \sectionmark{#1}}

\invisiblesection{Bibliography}
\bibliography{CookeKartaltepe2019bib}

\end{document}